\documentclass{aa}  

\usepackage{graphicx}
\usepackage{txfonts}
\usepackage[allcolors=blue]{hyperref}

\def\Msun{M_{\sun}}
\def\Lsun{L_{\sun}}

\def\Rstar{R_{\star}}

\begin{document} 

   \title{Dynamic atmospheres and winds of cool luminous giants}

   \subtitle{II. Gradual Fe enrichment of wind-driving silicate grains}

   \author{S. Höfner
          \inst{1}
          \and
          S. Bladh\inst{1}
          \and
          B. Aringer\inst{2,3}
          \and 
          K. Eriksson\inst{1}
          }

   \institute{Theoretical Astrophysics, Department of Physics and Astronomy, Uppsala University, 
              Box 516, 751 20 Uppsala, Sweden\\
              \email{susanne.hoefner@physics.uu.se}
         \and
         Dipartimento di Fisica e Astronomia Galileo Galilei,
         Universit\`a di Padova, Vicolo dell'Osservatorio 3, I-35122 Padova, Italy
         \and
         Department of Astrophysics, Univ.\ of Vienna,
         T\"urkenschanzstra{\ss}e 17, A-1180 Wien, Austria
             }

   \date{Received ...; accepted ...}

  \abstract
   {The winds observed around  asymptotic giant branch (AGB) stars are generally attributed to radiation pressure on dust formed in the extended dynamical atmospheres of these long-period variables. The composition of wind-driving grains is affected by a feedback between their optical properties and the resulting heating due to stellar radiation.
   }
   {We explore the gradual Fe enrichment of wind-driving silicate grains in M-type AGB stars to derive typical values for Fe/Mg and to test the effects on wind properties and synthetic spectra. 
   }
   {We present new radiation-hydrodynamical DARWIN models that allow for the growth of silicate grains with a variable Fe/Mg ratio and predict mass-loss rates, wind velocities, and grain properties. Synthetic spectra and other observables are computed {\em a posteriori} with the COMA code.
   }
   {The self-regulating feedback between grain composition and radiative heating, in combination with quickly falling densities in the stellar wind, leads to low values of Fe/Mg, typically a few percent. Nevertheless, the new models show distinct silicate features around 10 and 18 microns. Fe enrichment affects visual and near-IR photometry moderately, and the new DARWIN models agree well with observations in $(J-K)$ versus $(V-K)$ and {\em Spitzer} color-color diagrams. The enrichment of the silicate dust with Fe is a secondary process, taking place in the stellar wind on the surface of large Fe-free grains that have initiated the outflow. Therefore, the mass-loss rates are basically unaffected, while the wind velocities tend to be slightly higher than in corresponding models with Fe-free silicate dust. 
   }
   {The gradual Fe enrichment of silicate grains in the inner wind region should produce signatures observable in mid-IR spectro-interferometrical measurements. Mass-loss rates derived from existing DARWIN models, based on Fe-free silicates, can be applied to stellar evolution models since the mass-loss rates are not significantly affected by the inclusion of Fe in the silicate grains. 
   }


   \maketitle
%

\section{Introduction}

Dust grains forming around asymptotic giant branch (AGB) stars have a strong impact on the appearance and further evolution of these cool luminous giants. When interacting with stellar photons, the solid particles gain outward-directed momentum, which is transferred to the surrounding gas by collisions, leading to an outflow of gas and dust from the stellar atmosphere. Over time, the dust-driven wind of an AGB star severely reduces the stellar mass and builds up a massive circumstellar envelope that can be studied with a wide range of observational techniques 
\citep[for a recent review, see][]{hoefolof18}.  

Magnesium-iron silicates (olivine- and pyroxene-type materials) seem to be obvious candidates for wind-driving dust species in M-type AGB stars given the relatively high abundances of their constituent elements (Si, Mg, Fe, and O) and the prominence of silicate features in mid-IR spectra of circumstellar dust shells 
\citep[][]{woolney69,dorsetal10,molsetal10}.
Based on the comparable abundances of Mg and Fe at solar metallicities, it might be expected that they condense into grains in similar amounts. However, when taking into account radiative effects, a subtler picture emerges: The optical properties of silicates depend strongly on their Fe/Mg ratio, introducing a bias toward low Fe/Mg values in the close stellar vicinity. The inclusion of Fe atoms makes silicates very opaque at visual and near-IR wavelengths, and the resulting absorption of stellar photons leads to strong radiative heating
\citep[see, e.g.,][]{dorsetal95,jaegetal03,woit06,zeidetal11}. 
Fe-free olivine- and pyroxene-type silicates, on the other hand, are very transparent at visual and near-IR wavelengths, where the stellar flux has its maximum. They will get heated less by stellar radiation than their Fe-bearing counterparts, and they can exist closer to the stellar surface, which is critical for triggering a stellar wind \citep{hoef08}. 

While a low absorption efficiency is favorable in terms of grain temperature, there is a drawback for radiation pressure. In order to produce sufficient radiative acceleration for driving a wind, near-transparent grains have to gain outward momentum by efficiently scattering stellar photons. To maximize the impact of scattering on radiative pressure, the grains have to be of sizes that are comparable to the wavelengths near the stellar flux maximum (i.e., grain radii in the range of 0.1 -- 1 $\mu$m). When \citet{hoef08} introduced detailed AGB wind models based on this driving mechanism, the idea that grains would grow that large around AGB stars was controversial. Since then, however, dust grains of such sizes have been detected in the close vicinity of several nearby AGB stars \citep[e.g.,][]{norretal12,ohnaetal17}.

In earlier papers we demonstrated that radiation-hydrodynamical models of winds driven by photon scattering on Mg$_2$SiO$_4$ grains produce realistic mass-loss rates and wind velocities, as well as visual and near-IR spectra and photometry in good agreement with observations \citep{hoef08,bladetal13,bladetal15,bladetal19}. 
In the first paper of this series \citep{hoefetal16}, we also showed that the formation of composite grains, with a mantle of Mg$_2$SiO$_4$ condensing on a core of Al$_2$O$_3$, speeds up grain growth to the critical size range and may lead to even better agreement with visual and near-IR photometry. 

Here, we present new DARWIN models that allow for the growth of silicate grains with a variable Fe/Mg ratio. The grains start out with an Fe-free composition when first forming close to the star. Allowing for the inclusion of Fe atoms into the growing silicate grains, we can determine if the radiative acceleration of the wind is affected by an increased absorption efficiency. We can also address a known issue of our previous DARWIN models: a lack of discernible mid-IR features due to the low temperature of pure Mg$_2$SiO$_4$ grains \cite[see][]{bladetal15,bladetal17}. 

In Sect.~\ref{s_method} we describe the updated version of the DARWIN code and the underlying physical assumptions. In Sects.~\ref{s_results} and \ref{s_discussion} we present the new models and compare them to previous models and observations. The conclusions are summarized in Sect.~\ref{s_concl}.


\section{Upgraded DARWIN models}\label{s_method}

The models presented in this paper were computed with a new version of the DARWIN code. 
The upgraded treatment of silicate dust formation is discussed in Sect.~\ref{s_oli} and the corresponding modifications regarding the optical properties of the dust grains are described in Sect.~\ref{s_tdust}.
The general features of the code are described in detail in \citet[][Paper I]{hoefetal16}, and we only give a short overview below. 

\subsection{DARWIN models: General properties}\label{s_rhd}

The variable radial structure of the stellar atmosphere and wind is determined by the coupled system of gas dynamics, radiative processes, and nonequilibrium dust formation, describing the dependence of velocities, densities, temperatures and dust properties on time and distance from the stellar center. 
The spherically symmetric models cover a region with an inner boundary below the stellar photosphere but above the driving zone of the pulsations. The effects of stellar pulsation on the atmosphere are simulated by temporal variations in gas velocity and luminosity at the inner boundary of the model.
The computations start with a hydrostatic, dust-free, atmospheric structure corresponding to the fundamental parameters of the star. Pulsation effects are introduced gradually by increasing the amplitude up to the full value over typically tens of periods and the simulations are run for several hundred pulsation periods to avoid transient effects.

Mass-loss rates, wind velocities and dust properties are a direct output of the DARWIN code. Spectra, light curves and other synthetic observables are computed {\it a posteriori} from snapshots of the radial structures using the COMA code \citep{arinetal09,arinetal16}. 

\subsection{Growth and composition of silicate grains}\label{s_oli}

Olivine-type silicates (i.e., Mg$_{2x}$Fe$_{2(1-x)}$SiO$_4$ with $0<x<1$) can be regarded as a solid solution of Mg$_2$SiO$_4$ and Fe$_2$SiO$_4$, with the value of $x$ being set by the conditions in the grain growth region. We describe the formation of Mg$_2$SiO$_4$ from abundant molecules in the gas phase according to the net reaction
\begin{equation}\label{e_path_ol}
  {\rm 2 \, Mg + SiO + 3 \, H_2 O }  \,\,  \longrightarrow  \,\,  {\rm Mg_2 SiO_4 + 3 \, H_2 } \, ,
\end{equation}
and the formation of Fe$_2$SiO$_4$ by a second net reaction
\begin{equation}\label{e_path_ol2}
  {\rm 2 \, Fe + SiO + 3 \, H_2 O }  \,\,  \longrightarrow  \,\,  {\rm Fe_2 SiO_4 + 3 \, H_2 } \, .
\end{equation}
Both materials are assumed to condense onto composite grains, allowing for varying Fe/Mg ratios in the silicates that depend on the growth history of the dust particles. 

Initially, close to the star, silicate grains will mostly grow by the first of the two reactions, since the inclusion of Fe introduces extra radiative heating. As the material is transported outward in the stellar wind, and the radiative flux from the star weakens with distance, it can be expected that thin mantles of Fe-bearing silicates form on top of Fe-free wind-driving grains, before grain growth effectively comes to a halt due to falling densities.  

The changing size and composition (Fe/Mg) of the olivine-type silicate grains in the atmosphere and wind are described by two equations, specifying the net rates at which monomers (i.e., the basic building blocks Mg$_2$SiO$_4$ and Fe$_2$SiO$_4$) are added to or removed from the grain surface. The generic form of these two equations is 
\begin{equation}\label{e_rate_ol}
  \frac{d N}{d t} 
    = 4 \pi \, a^2_{\rm oli} \, \left[  J^{\rm gr}  - J^{\rm dec} \right] 
,\end{equation}
where $N$ is the number of monomers of a specific type (Mg$_2$SiO$_4$ or Fe$_2$SiO$_4$) in the composite grain, $4 \pi \, a^2_{\rm oli}$ is the grain surface area ($a_{\rm oli}$ = grain radius), and $J^{\rm gr}$ and $J^{\rm dec}$ denote the rates per surface area at which monomers are added or lost by reactions with the surrounding gas and thermal evaporation. 

As discussed in Paper I \citep{hoefetal16}, for Mg$_2$SiO$_4$ the growth and decomposition rates per surface area can be formulated as 
\begin{eqnarray}
  J^{\rm gr}   & = & \frac{1}{2} \, \alpha_{\rm Mg} v_{\rm Mg} n_{\rm Mg} 
                                         \label{e_jgr_ol_mg} \\
  J^{\rm dec}  & = & \frac{1}{2} \, \alpha_{\rm Mg} v_{\rm Mg} \frac{p_{\rm v , Mg}}{k T_g} \sqrt{\frac{T_g}{T_d}}
,\end{eqnarray}
respectively (accounting for different gas and dust temperatures, $T_g$ and $T_d$), for a star with a solar element mixture. 
The root mean square thermal velocity of the Mg atoms in the gas phase is given by $v_{\rm Mg} = \sqrt{k T_g / 2 \pi m_{\rm Mg}}$, the symbol $n_{\rm Mg}$ denotes the number density of Mg atoms in the gas phase (taking depletion of Mg due condensation into account) and $\alpha_{\rm Mg}$ is the sticking coefficient. The quantity $p_{\rm v , Mg}$ appearing in the decomposition rate represents the partial pressure of Mg in chemical equilibrium between the gas phase and Mg$_2$SiO$_4$, according to the net reaction in Eq.\,(\ref{e_path_ol}) and its reverse process. In a solar element mixture the abundances of Si and Mg are comparable, and the abundance of SiO will be determined by the abundance of Si in the gas phase. Since 2~Mg atoms are required for each SiO to complete a monomer (see Eq.~\ref{e_path_ol}), Mg will be the limiting factor for the condensation of Mg$_2$SiO$_4$. 

For the formation of Fe$_2$SiO$_4$, the critical constituent in a solar mixture will be Fe, since two~Fe atoms are required for each SiO to complete a monomer (see Eq.~\ref{e_path_ol2}), and the abundances of Fe and Si are comparable. Consequently, the corresponding growth and decomposition rates for Fe$_2$SiO$_4$ can be expressed as 
\begin{eqnarray}
  J^{\rm gr}   & = & \frac{1}{2} \, \alpha_{\rm Fe} v_{\rm Fe} n_{\rm Fe} 
                                         \label{e_jgr_ol_fe} \\
  J^{\rm dec}  & = & \frac{1}{2} \, \alpha_{\rm Fe} v_{\rm Fe} \frac{p_{\rm v , Fe}}{k T_g} \sqrt{\frac{T_g}{T_d}}
,\end{eqnarray}
where the root mean square thermal velocity of the Fe atoms in the gas phase is given by $v_{\rm Fe} = \sqrt{k T_g / 2 \pi m_{\rm Fe}}$, the symbol $n_{\rm Fe}$ denotes the number density of Fe atoms in the gas phase (taking depletion of Fe due condensation into account), $\alpha_{\rm Fe}$ is the sticking coefficient, and $p_{\rm v , Fe}$ represents the partial pressure of Fe in chemical equilibrium between the gas phase and Fe$_2$SiO$_4$, according to the net reaction in Eq.\,(\ref{e_path_ol2}) and its reverse process. 

In the models presented here, $p_{\rm v , Mg}$ and $p_{\rm v , Fe}$ were calculated based on the data for free energies by \citet[]{sharhueb90}. Regarding sticking coefficients, we argued in Paper I that $\alpha_{\rm Mg} = 1$ is a reasonable assumption, considering that Mg$_2$SiO$_4$ dust is much cooler than the surrounding gas. For Fe$_2$SiO$_4$, on the other hand, considerable radiative heating has to be expected. Nevertheless, for simplicity, we assumed $\alpha_{\rm Fe} = 1$. This corresponds to very favorable growth conditions, and probably gives an upper limit to Fe/Mg values of grains in the atmosphere and inner wind region (further discussed in Sect.~\ref{s_discussion}). 

The radius $a_{\rm oli}$ of a composite grain, used in the rate equations given above, is derived from the total grain volume. The latter is determined by adding up the numbers of the two types of monomers forming the grain, multiplied with the respective nominal volumes of the monomers. The monomer volume is defined as $V_{\rm mon} = A_{\rm mon} \, m_{\rm H} / \rho_{\rm bulk}$, where $A_{\rm mon}$ is the atomic weight of the monomer, $\rho_{\rm bulk}$ the bulk density of the dust material, and $m_{\rm H}$ the hydrogen mass. 
For Mg$_2$SiO$_4$ we used $A_{\rm mon} = 140$ and a bulk density of $\rho_{\rm bulk} = 3.27$ [g/cm$^3$]. The corresponding values for Fe$_2$SiO$_4$ were $A_{\rm mon} = 204$ and $\rho_{\rm bulk} = 4.40$ [g/cm$^3$].

It should be noted here that the equations given above describe how the sizes of existing grains change by condensation onto or evaporation from the grain surface. The equations do not include the process of nucleation (i.e., the formation of new solid particles out of the gas phase). As discussed in Paper I, grain nucleation in M-type AGB stars is an unsolved problem. The nature of the initial condensate is still strongly debated, and no applicable nucleation rates exist at present. Therefore, we treated the abundance of seed particles $n_d/n_{\rm H}$ as a free parameter, which was varied over a wide range, while keeping all other parameters fixed, to demonstrate its influence on the results (see Sect.~\ref{s_results}). 

\subsection{Optical properties and grain temperature}\label{s_tdust}

The equations discussed in the previous section give basic properties of olivine-type silicate grains (size, Fe/Mg) throughout the region covered by the models. Since the dust grains are not necessarily small compared to all wavelengths of interest, however, we need to know more about their internal structure for computing their optical properties and their resulting temperature when exposed to the stellar radiation field. 

To begin with, it is reasonable to assume that the innermost part of the silicate grains is Fe-free. From a purely chemical point of view, Fe$_2$SiO$_4$ would start forming later than Mg$_2$SiO$_4$ (further away from the star), due to a somewhat lower condensation temperature. 
Additionally, being exposed to the stellar radiation, Fe-bearing grains are heated more strongly, due to their higher absorption efficiency in the near-IR. Consequently, an even larger distance from the stellar surface is required to reach grain temperatures below the condensation threshold of Fe-bearing silicates. It can therefore be expected that a significant enrichment of the grains with Fe will only take place after the outflow has been initiated, which typically requires Mg$_2$SiO$_4$ grains with radii well above 0.1 micron. Once both growth reactions (Eq.~\ref{e_path_ol} and \ref{e_path_ol2}) become efficient, a mixed mantle consisting of Mg$_2$SiO$_4$ and Fe$_2$SiO$_4$ may form on top of the Fe-free inner core. 

The changing value of Fe/Mg from the center to the surface of a grain is a record of its growth history and can, in principle, be derived when solving the equations given in Sect.~\ref{s_oli}. 
In practice, this is easy in steady-state wind models (where distance from the star traces the temporal evolution), and a reasonable effort in time-dependent models based on co-moving frame formulations. The DARWIN code, however, uses a self-adaptive spatial grid (see Paper\,I, Appendix\,C), and keeping track of the grain growth history is a nontrivial task, beyond the scope of this study. 

For the purpose of computing optical properties of the olivine-type silicate grains, we assumed that the internal structure can be described by a simplified core-mantle structure. In the models presented here, we approximated the structure by a mantle of MgFeSiO$_4$ on top of an Mg$_2$SiO$_4$ core. The choice of $x=0.5$ (equal amounts of Mg and Fe) for the mantle material is based on estimates for the growth rates of Mg$_2$SiO$_4$ and Fe$_2$SiO$_4$ (see Appendix\,\ref{app_x05}). The relative sizes of core and mantle were derived from the Fe/Mg ratio of a grain, resulting from the rate equations discussed above. This two-layer approximation allowed us to use standard Mie theory when calculating the optical properties of olivine-type silicate grains with Fe/Mg and size determined by the equations given in Sect.~\ref{s_oli}. In particular, we used the {\em Fortran} subroutine DMiLay.f \cite[][]{toonacke81} to compute the absorption and scattering efficiencies as well as the mean scattering angle. From these quantities, we derived the grain temperature and the radiative pressure exerted on the grains (see Paper I for details). 

\begin{figure}
\centering
\includegraphics[width=\hsize]{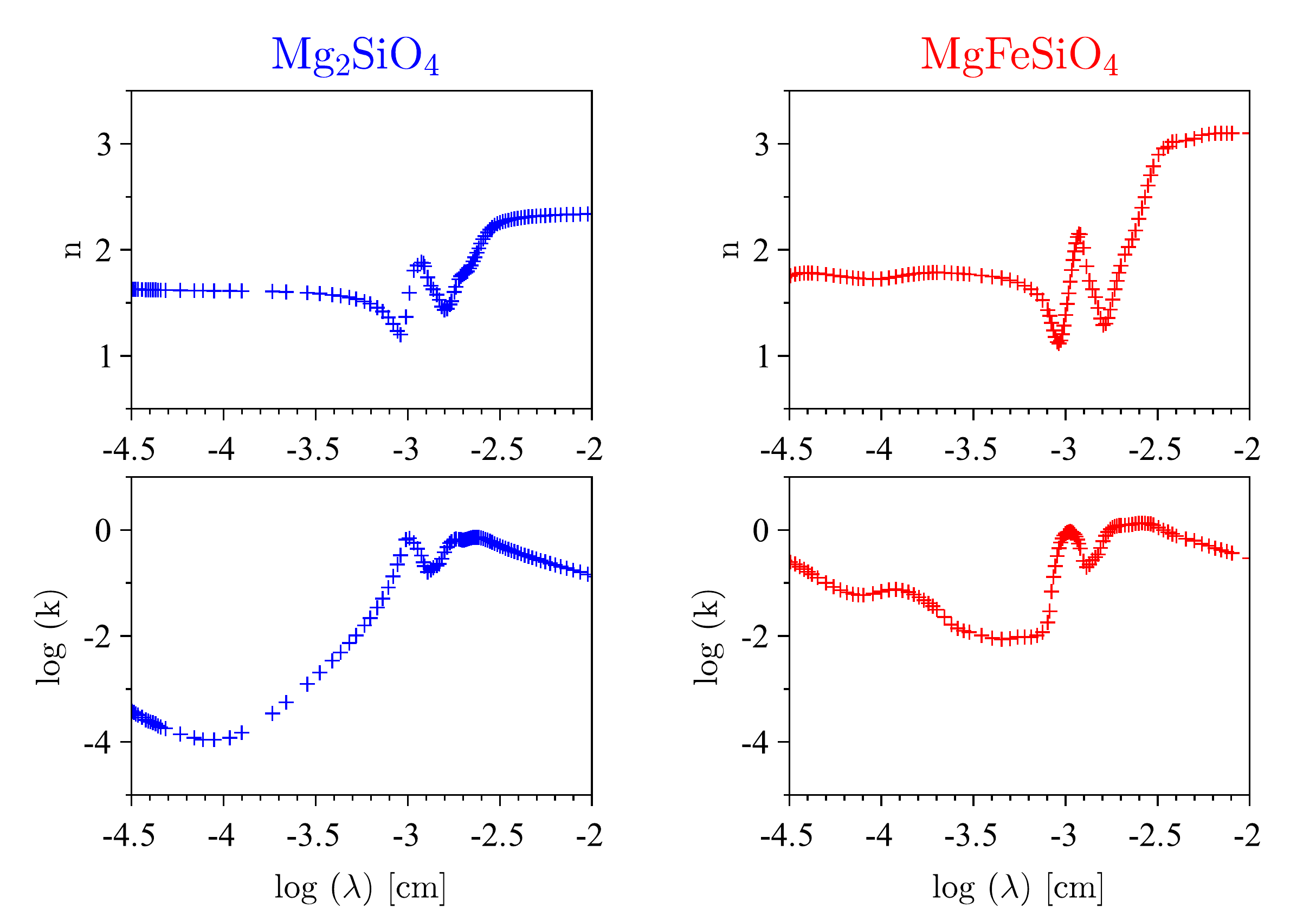}
     \caption{Refractive index data. 
                   {\em Left panels:} Mg$_2$SiO$_4$ \citep[][]{jaegetal03}.   
                   {\em Right panels:} MgFeSiO$_4$ \citep{dorsetal95}.
                   }
      \label{f_nk_sil}
\end{figure}

%
\begin{table*}
\caption{\label{t_mod}DARWIN models with outflows driven by the scattering of stellar photons on olivine-type silicate grains.}
\centering
\begin{tabular}{l|ccc|cccccccc}
\hline\hline
  & & & & & & \\
  model &  $n_d/n_{\rm H}$  &  $\Delta u_{\rm P}$ & $f_L$ & $\Delta M_{\rm bol}$ 
  & $\dot{M}$  & $u_{\rm ext}$  & $f_{\rm Si}$  & $f_{\rm Mg}$ & $f_{\rm Fe}$ & $a_{gr}$ & Fe/Mg \\
  name  &  & [km/s] &  & [mag]
  & [$\Msun$/yr] & [km/s] &  &  &  & [$\mu$m] & \\ 
  & & & & & & \\
\hline
  & & & & & & \\
  An115u3 / oli & $1.0 \cdot 10^{-15}$ & 3.0 & 2.0 & 0.54 
  & $1 \cdot 10^{-7}$ & 4 & 0.18 & 0.32 & 0.014 & 0.48 & 0.037 \\
  An315u3 / oli  & $3.0 \cdot 10^{-15}$ & 3.0 & 2.0 & 0.54
  & $3 \cdot 10^{-7}$ & 7 & 0.21 & 0.37 & 0.016 & 0.35 & 0.036 \\
  An114u3 / oli & $1.0 \cdot 10^{-14}$ & 3.0 & 2.0 & 0.54
  & $3 \cdot 10^{-7}$ & 8 & 0.28 & 0.50 & 0.026 & 0.26 & 0.043 \\
  An314u3 / oli & $3.0 \cdot 10^{-14}$ & 3.0 & 2.0 & 0.54
  & $4 \cdot 10^{-7}$ & 8 & 0.40 & 0.71 & 0.044 & 0.20 & 0.052 \\
  & & & & & & \\
  An115u4 / oli & $1.0 \cdot 10^{-15}$ & 4.0 & 2.0 & 0.73 
  & $3 \cdot 10^{-7}$ & 4 & 0.17 & 0.30 & 0.017 & 0.47 & 0.048 \\
  An315u4 / oli  & $3.0 \cdot 10^{-15}$ & 4.0 & 2.0 & 0.73 
  & $4 \cdot 10^{-7}$ & 6 & 0.20 & 0.36 & 0.017 & 0.34 & 0.040 \\
  An114u4 / oli & $1.0 \cdot 10^{-14}$ & 4.0 & 2.0 & 0.73 
  & $5 \cdot 10^{-7}$ & 9 & 0.28 & 0.50 & 0.026 & 0.26 & 0.043 \\
  An314u4 / oli & $3.0 \cdot 10^{-14}$ & 4.0 & 2.0 & 0.73 
  & $4 \cdot 10^{-7}$ & 8 & 0.39 & 0.69 & 0.043 & 0.20 & 0.052 \\
  & & & & & & \\
\hline
  & & & & & & \\
  Bn316u3 / oli & $3.0 \cdot 10^{-16}$ & 3.0 & 2.0 & 0.53 
  & $3 \cdot 10^{-7}$ & 3 & 0.14 & 0.24 & 0.012 & 0.65 & 0.041 \\
  Bn115u3 / oli & $1.0 \cdot 10^{-15}$ & 3.0 & 2.0 & 0.53 
  & $5 \cdot 10^{-7}$ & 6 & 0.14 & 0.24 & 0.015 & 0.44 & 0.052 \\
  Bn315u3 / oli & $3.0 \cdot 10^{-15}$ & 3.0 & 2.0 & 0.53 
  & $7 \cdot 10^{-7}$ & 8 & 0.18 & 0.32 & 0.019 & 0.33 & 0.050 \\
  Bn114u3 / oli & $1.0 \cdot 10^{-14}$ & 3.0 & 2.0 & 0.53 
  & $7 \cdot 10^{-7}$ & 10 & 0.25 & 0.45 & 0.031 & 0.25 & 0.058 \\
  Bn314u3 / oli & $3.0 \cdot 10^{-14}$ & 3.0 & 2.0 & 0.53 
  & $7 \cdot 10^{-7}$ & 10 & 0.36 & 0.64 & 0.051 & 0.20 & 0.067 \\
  & & & & & & \\
  Bn316u4 / oli & $3.0 \cdot 10^{-16}$ & 4.0 & 2.0 & 0.71 
  & $3 \cdot 10^{-7}$ & 3 & 0.13 & 0.22 & 0.013 & 0.64 & 0.049 \\
  Bn115u4 / oli & $1.0 \cdot 10^{-15}$ & 4.0 & 2.0 & 0.71 
  & $5 \cdot 10^{-7}$ & 5 & 0.14 & 0.24 & 0.014 & 0.44 & 0.048 \\
  Bn315u4 / oli & $3.0 \cdot 10^{-15}$ & 4.0 & 2.0 & 0.71 
  & $7 \cdot 10^{-7}$ & 8 & 0.18 & 0.32 & 0.019 & 0.33 & 0.050 \\
  Bn114u4 / oli & $1.0 \cdot 10^{-14}$ & 4.0 & 2.0 & 0.71 
  & $8 \cdot 10^{-7}$ & 10 & 0.25 & 0.45 & 0.031 & 0.25 & 0.058 \\
  Bn314u4 / oli & $3.0 \cdot 10^{-14}$ & 4.0 & 2.0 & 0.71 
  & $7 \cdot 10^{-7}$ & 10 & 0.36 & 0.64 & 0.051 & 0.20 & 0.067 \\
  & & & & & & \\
\hline
  & & & & & & \\
  M2n115u6 / oli & $1.0 \cdot 10^{-15}$ & 6.0 & 1.5 & 0.95 
  & $1 \cdot 10^{-6}$ & 6 & 0.20 & 0.36 & 0.019 & 0.49 & 0.045 \\
  M2n315u6 / oli & $3.0 \cdot 10^{-15}$ & 6.0 & 1.5 & 0.95 
  & $2 \cdot 10^{-6}$ & 11 & 0.24 & 0.43 & 0.026 & 0.37 & 0.051 \\
  M2n114u6 / oli & $1.0 \cdot 10^{-14}$ & 6.0 & 1.5 & 0.95 
  & $2 \cdot 10^{-6}$ & 16 & 0.35 & 0.62 & 0.044 & 0.28 & 0.060 \\
  & & & & & & \\
\hline
\end{tabular}
\tablefoot{This table gives an overview of input parameters and resulting properties of the models: 
$n_d/n_{\rm H}$ is the seed particle abundance, and $\Delta u_{\rm P}$ is the velocity amplitude at the inner boundary. This amplitude, in combination with the parameter $f_L$ (see Paper I, Appendix B), determines the luminosity amplitude and the corresponding bolometric amplitude $\Delta M_{\rm bol}$. The resulting wind and dust properties listed here are temporal means of the mass-loss rate $\dot{M}$, the wind velocity $u_{\rm ext}$, the fraction of Si condensed into grains $f_{\rm Si}$, the fraction of Mg condensed into grains $f_{\rm Mg}$, the fraction of Fe condensed into grains $f_{\rm Fe}$, the grain radius $a_{gr}$, and the ratio of Mg and Fe in the grains, all taken at the outer boundary of the models (located at about 20-30 stellar radii). 
The model names are constructed in the following way: A, B, or M2 represents a combination of stellar parameters (see Sect.~\ref{s_results}), the letter n followed by a 3-digit number stands for the seed particle abundance (i.e., n315 for $n_d/n_{\rm H} = 3 \cdot 10^{-15}$, n114 for $n_d/n_{\rm H} = 1 \cdot 10^{-14}$, etc.) and the letter u followed by a number (3, 4, 6) represents the velocity amplitude at the inner boundary (km/s). We note that the parameter $\Delta u_{\rm P}$ is the half-amplitude of the sinusoidal velocity variation (see Paper I, Appendix\,B), while $\Delta M_{\rm bol}$ listed in this table represents the full bolometric amplitude (minimum to maximum). 
}
\end{table*}

The refractive index data for Mg$_2$SiO$_4$ and MgFeSiO$_4$ used in the models were taken from \citet[][]{jaegetal03} and \citet[][]{dorsetal95}, respectively. The two $n$ and $k$ data sets are plotted in Fig.~\ref{f_nk_sil}. Comparing the values of the imaginary part of the refractive index $k$ (which defines true absorption, and thereby radiative heating), we see that the values for MgFeSiO$_4$ are about three orders of magnitude higher than for Mg$_2$SiO$_4$ in the region around 1\,$\mu$m, where the stellar flux peaks. Strong radiative heating, compared to Fe-free grains, can therefore be expected even for thin mantles of Fe-bearing silicates. In the mid-IR region, the values of $k$ are much more comparable, and both species show local maxima around 10 and 20 microns, which lead to the characteristic emission features observed in AGB star spectra.


\begin{figure*}
\centering
\includegraphics[width=18.2 cm]{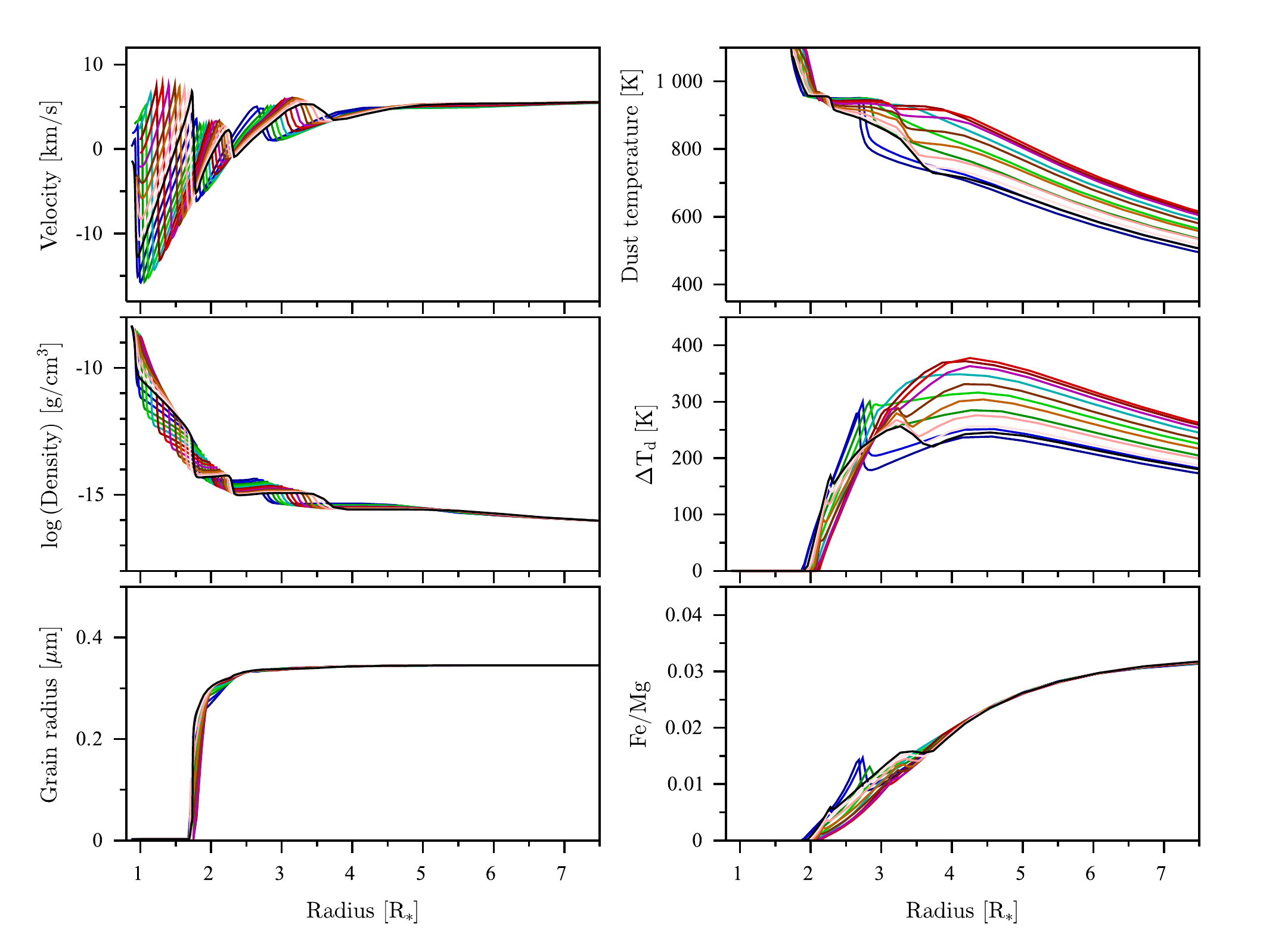}
     \caption{Time-dependent radial structure of model An315u3, zoomed in on the dust formation region 
     (snapshots of 13 phases during a pulsation cycle). 
     {\em Left, top to bottom}: Flow velocity, gas density, grain radius. 
     {\em Right, top to bottom}: Temperature of the Fe-bearing silicate grains, difference in grain temperature with and without Fe, and the Fe/Mg ratio in the dust grains; see text for details. The first snapshot, shown in dark blue, corresponds to a near-minimum phase, green colors represent the ascending part of the bolometric light curve, the red curves show phases close to the luminosity maximum, and the remaining colors represent the descending part of the bolometric light curve, ending with the black curve. It should be noted that only the innermost, dust-free parts of the model structures show periodic variations that repeat every pulsation cycle (with the final black curve close to the initial dark blue curve in the velocity and density plots), while grain growth and wind acceleration are governed by other timescales. 
     }
      \label{f_struct}
\end{figure*}

\section{Model parameters and results}\label{s_results}

We used several sets of stellar parameters for the hydrostatic initial models (combinations of stellar mass $M_{\ast}$, luminosity $L_{\ast}$, and effective temperature $T_{\ast}$), labeled A, B, and M2 for easy reference (see Table\,\ref{t_par}).
\begin{table}
\caption{\label{t_par} Stellar parameters of the hydrostatic initial models (mass $M_{\ast}$, luminosity $L_{\ast}$, effective temperature $T_{\ast}$) and pulsation period $P$. 
        }
\begin{center}
\begin{tabular}{l|cccc}
\hline\hline
      & $M_{\ast}$  [$\Msun$] & $L_{\ast}$ [$\Lsun$] & $T_{\ast}$ [K] & $P$ [d] \\
\hline
   A & 1.0 & 5000 & 2800 & 310 \\
   B & 1.0 & 7000 & 2700 & 390 \\
   M2 & 1.5 & 7000 & 2600 & 490 \\
\hline
\end{tabular}
\end{center}
\end{table}
Parameter sets A and B, which result in low to moderate mass-loss rates and wind velocities, were also used in Paper I. Here we added parameter set M2, leading to stronger winds. An earlier version of this model (with Fe-free silicate dust) was used by \citet{liljetal17} to study the effects of luminosity variations on wind properties and synthetic observables. 

In addition to the stellar parameters ($M_{\ast}$, $L_{\ast}$, $T_{\ast}$) and the pulsation period $P$ (see Table\,\ref{t_par}), we also have to specify several other parameters for each model: The pulsation amplitude is defined by the velocity amplitude $\Delta u_{\rm P}$ at the inner boundary and the luminosity amplitude factor $f_L$. For the stellar parameter sets A and B we applied different velocity amplitudes at the inner boundary, as in Paper I. For each configuration of stellar and pulsation parameters, the seed particle abundance $n_d/n_{\rm H}$ (i.e., the ratio of the number densities of dust grains and H nuclei) was varied over a range of values (see Table~\ref{t_mod}).
At present, grain nucleation (i.e., the formation of seed particles directly out of the gas phase) is an unsolved problem for M-type AGB stars (see Sect.\,\ref{s_oli}). However, it is reasonable to assume that nucleation rates depend on local densities and temperatures, which are both highly variable quantities. Using a range of seed particle abundances for fixed stellar and pulsation parameters gives an indication of how changes in nucleation rates may affect model results.  

In the following, we first discuss the mechanism that defines Fe/Mg in the silicate grains, illustrated in detail with a selected model. Then we proceed to a discussion of wind and dust properties for the whole set of models. Finally, we compare the new models to observations of AGB stars. 

\subsection{Fe/Mg in silicates: Self-regulation via grain temperature}\label{s_femg}

Figure~\ref{f_struct} shows the time-dependent radial structure of a typical model, zooming in on the region where dust formation and wind acceleration take place. The inner part of the atmosphere (below about 2 stellar radii) is dominated by pulsation-induced shock waves, which are visible as steep changes in velocity in the top left panel. At a shock front, outward-moving gas meets infalling material from a previous pulsation cycle that has not been accelerated beyond the escape velocity. The density in the inner atmosphere is declining strongly with distance from the star, as seen in the middle left panel, and shows steep jumps where the gas is compressed by the propagating shock waves. The dense wakes of shocks may become regions of efficient dust formation, once the temperature has dropped below the condensation threshold. 

Around 1.6 stellar radii, dust condensation starts, and the grains grow rapidly in size (bottom left panel). Initially, the grains are Fe-free and very transparent at visual and near-IR wavelengths. When they reach the critical size regime where photon scattering becomes efficient, radiation pressure triggers an outflow (positive values of the velocities, top left). As the flow approaches its terminal velocity, the mean density profile changes from the steep atmospheric decline toward $\rho \propto 1/r^2$, characteristic of a wind with constant velocity. Grain growth slows down critically in the wind due to rapidly falling densities as the material is driven away from the star, leaving a considerable fraction of Si, Mg and Fe in the gas phase. 

\begin{figure}
\centering
\includegraphics[width=\hsize]{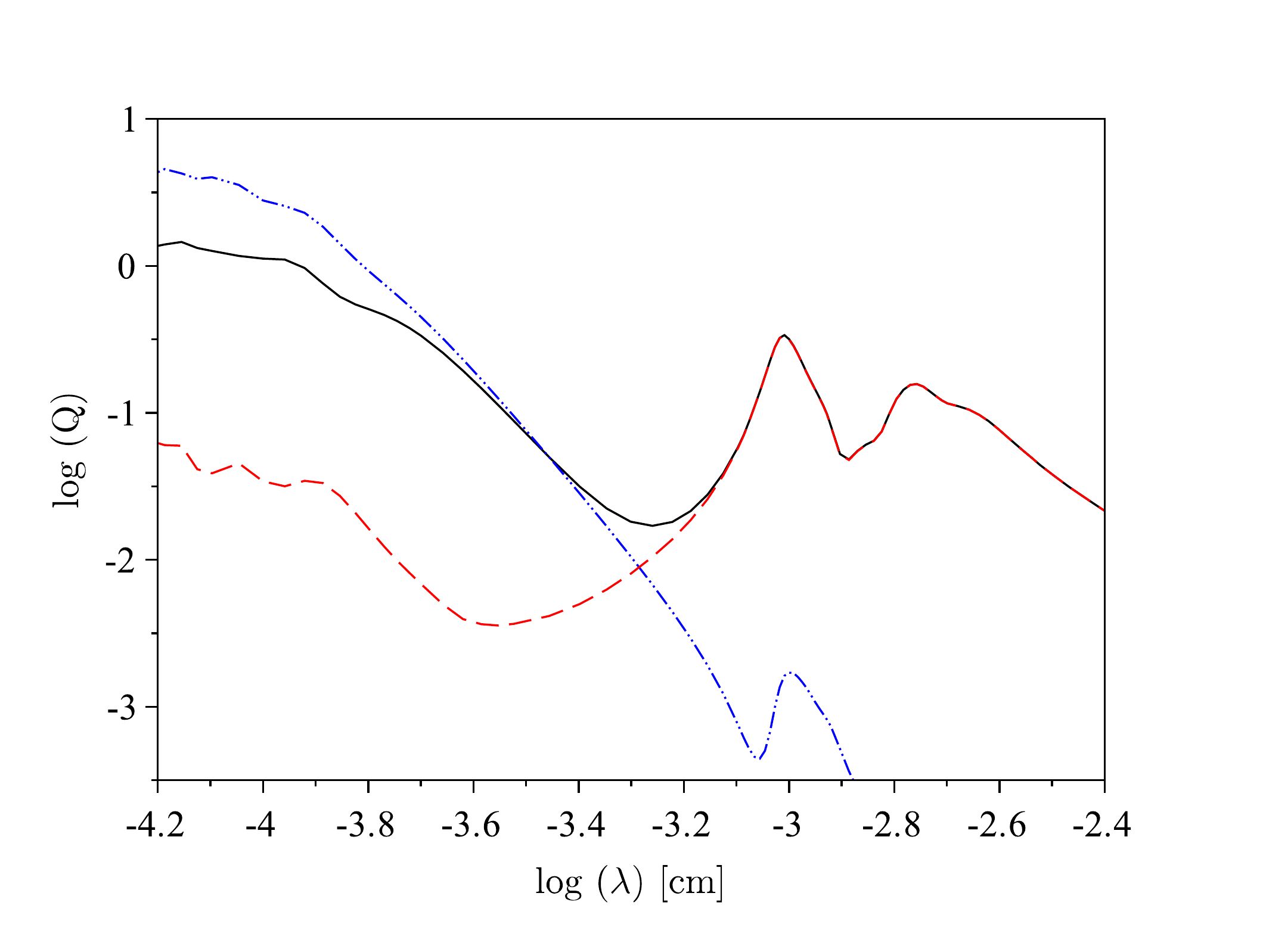}
\includegraphics[width=\hsize]{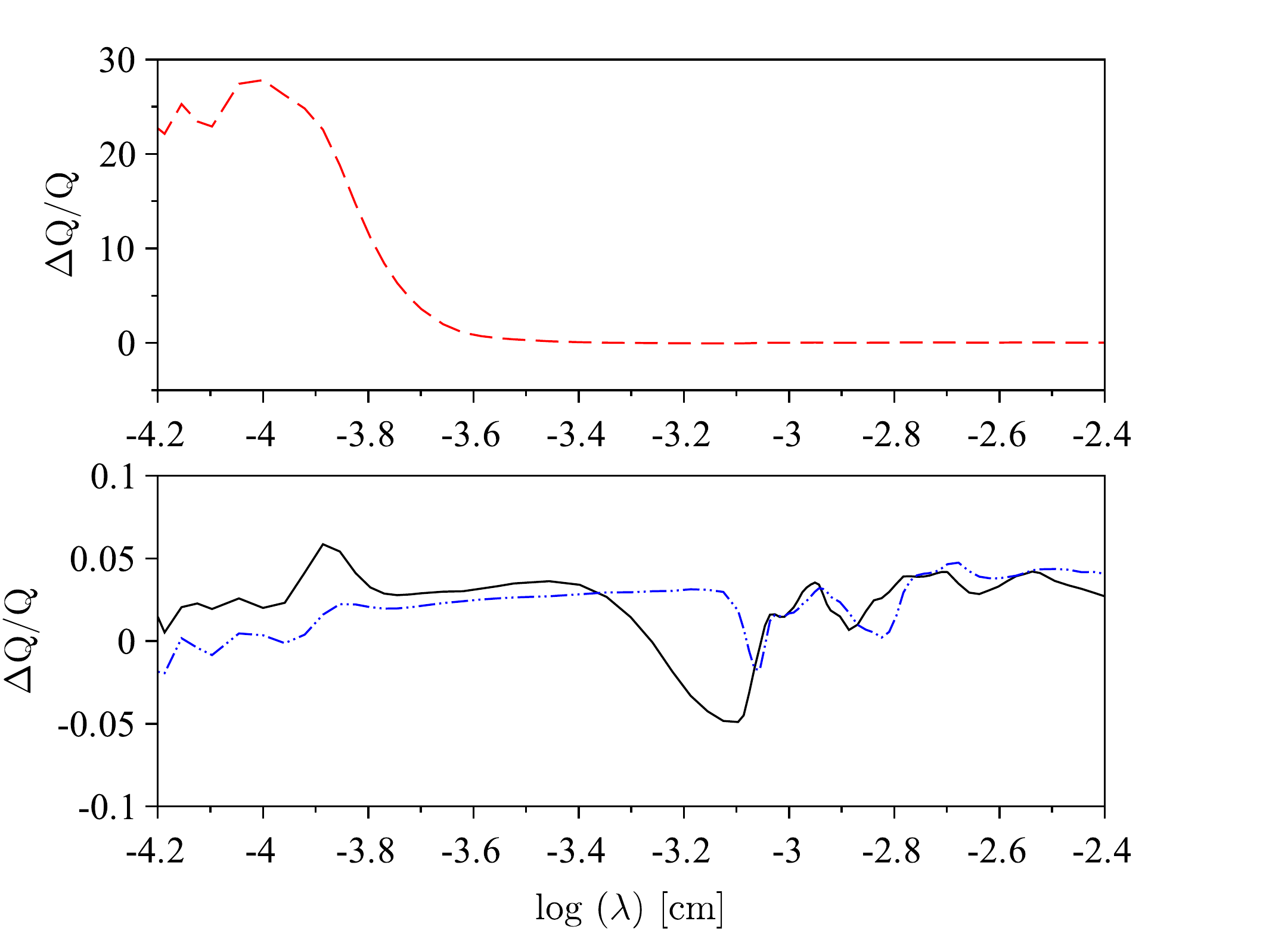}
     \caption{Optical properties of a silicate grain with a total radius of 0.35\,$\mu$m, consisting of an Mg$_2$SiO$_4$ core and an MgFeSiO$_4$ mantle corresponding to 2 percent of the grain radius. 
     {\em Top panel}: 
     Absorption efficiency (red dashed line), scattering efficiency (blue dash-triple-dotted line), and the resulting efficiency factor for radiative pressure (black line). 
     {\em Middle panel:} Relative difference of the absorption efficiency for the composite grain plotted in the top panel compared to a grain consisting entirely of Mg$_2$SiO$_4$. 
     {\em Bottom panel:} Same as the middle panel, but for the scattering efficiency (blue dash-triple-dotted line) and the efficiency factor for radiative pressure (black line). We note the different scales. The absorption efficiency of the Fe-enriched grain is almost 30 times higher around 1\,$\mu$m (near the stellar flux maximum) than that of the Fe-free grain, leading to significantly stronger radiative heating. The radiative pressure efficiency factor, dominated by the almost unchanged scattering efficiency (weighted with the scattering angle; see Paper I), on the other hand, is only slightly higher for the Fe-bearing composite grain in this spectral region, explaining the moderate effect on the wind velocity.  
             } 
      \label{f_FeMg_qg}
\end{figure}

In the wind, where the stellar flux and the resulting radiative heating of the dust grains decrease with distance from the star, dust particles can be more opaque without getting destroyed. As shown in the bottom right panel of Fig.\,\ref{f_struct}, the Fe/Mg ratio in the grains indeed increases as they move away from the star. However, the Fe enrichment is limited: even a small amount of Fe in the grains leads to substantial heating by absorption of stellar photons. The difference between the actual temperature of the composite silicate grains, including both Fe and Mg, and the hypothetical temperature of grains consisting entirely of Mg$_2$SiO$_4$ is shown in the middle right panel. Fe/Mg values of a few percent make the composite grains about $200 - 400$\,K warmer than their Fe-free counterparts. The top values are close to the sublimation temperature of the grains, apparent as a plateau in the top right panel, showing grain temperatures. 

In summary, we see the following picture emerging: Like in earlier models, the outflow is initiated by Fe-free silicate grains, as soon as they have grown to sizes where radiative acceleration due to photon scattering becomes efficient. As the stellar flux, which heats the dust, drops with increasing distance, a gradual enrichment of the grains with Fe begins. The value of Fe/Mg in the grains results from a self-regulating feedback between the growth rates (Eqs. \ref{e_path_ol} and \ref{e_path_ol2}) and composition-dependent radiative heating, which sets an upper limit to the Fe-content. The rising Fe/Mg is accompanied by an increase in absorption, and thereby higher radiative pressure on the grains (see Fig.~\ref{f_FeMg_qg}). This leads to a somewhat higher wind velocity than in a model with Fe-free silicate dust. The mass-loss rate, on the other hand, is basically unaffected. This is consistent with expectations, as the wind is initiated by Fe-free dust, and the Fe-bearing mantles of the grains are formed in the outflow.    

\begin{figure}
\centering
\includegraphics[width=\hsize]{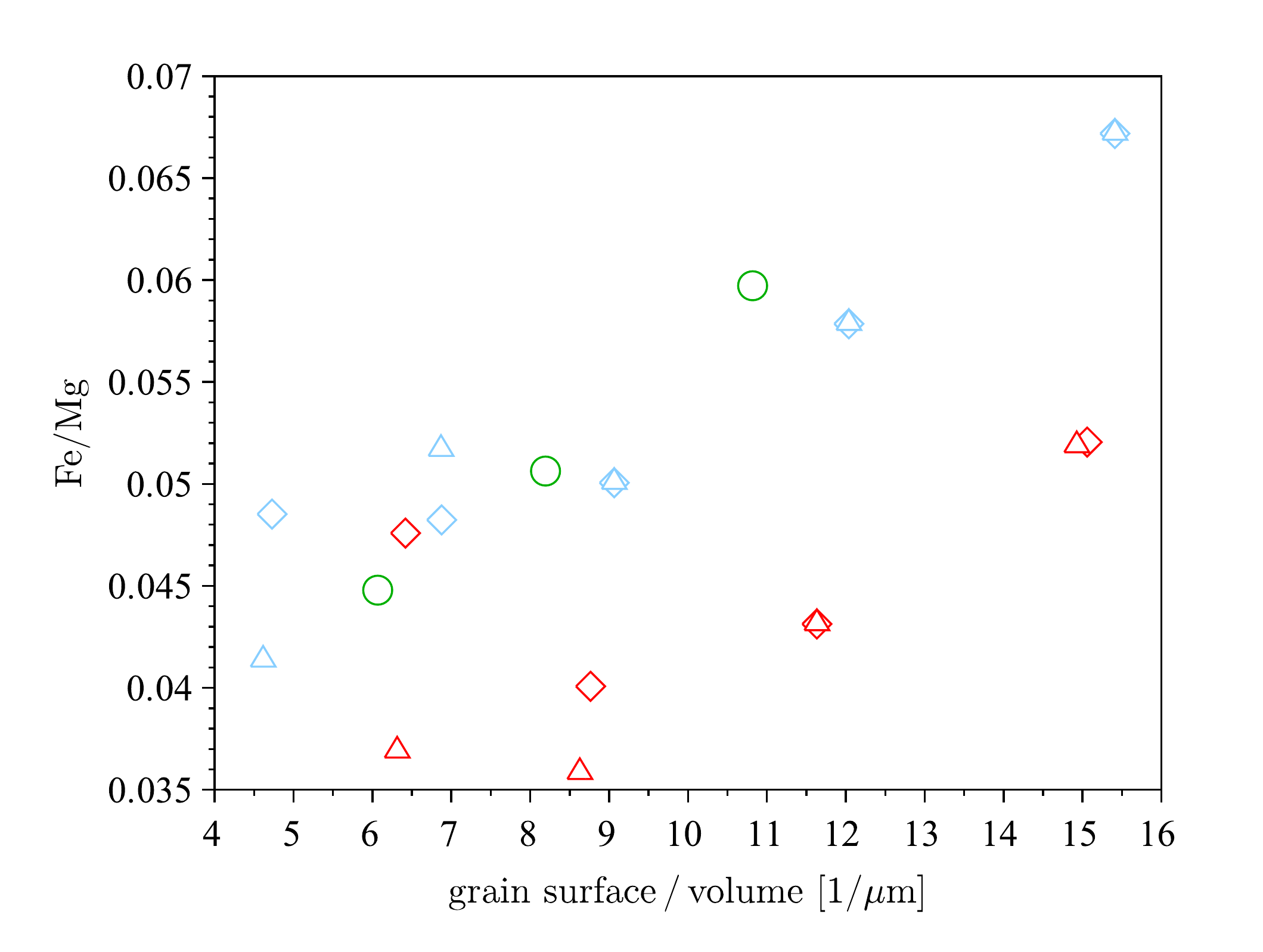}
     \caption{Fe/Mg versus surface/volume ratio of the grains. Red symbols mark models of series A and blue symbols models of series B (triangles and diamonds indicate pulsation amplitudes of 3 and 4 km/s, 
respectively); models of series M2 are shown as green circles (see Table~\ref{t_mod} for details). 
             }
      \label{f_FeMg_s2v}
\end{figure}

\begin{figure}
\centering
\includegraphics[width=\hsize]{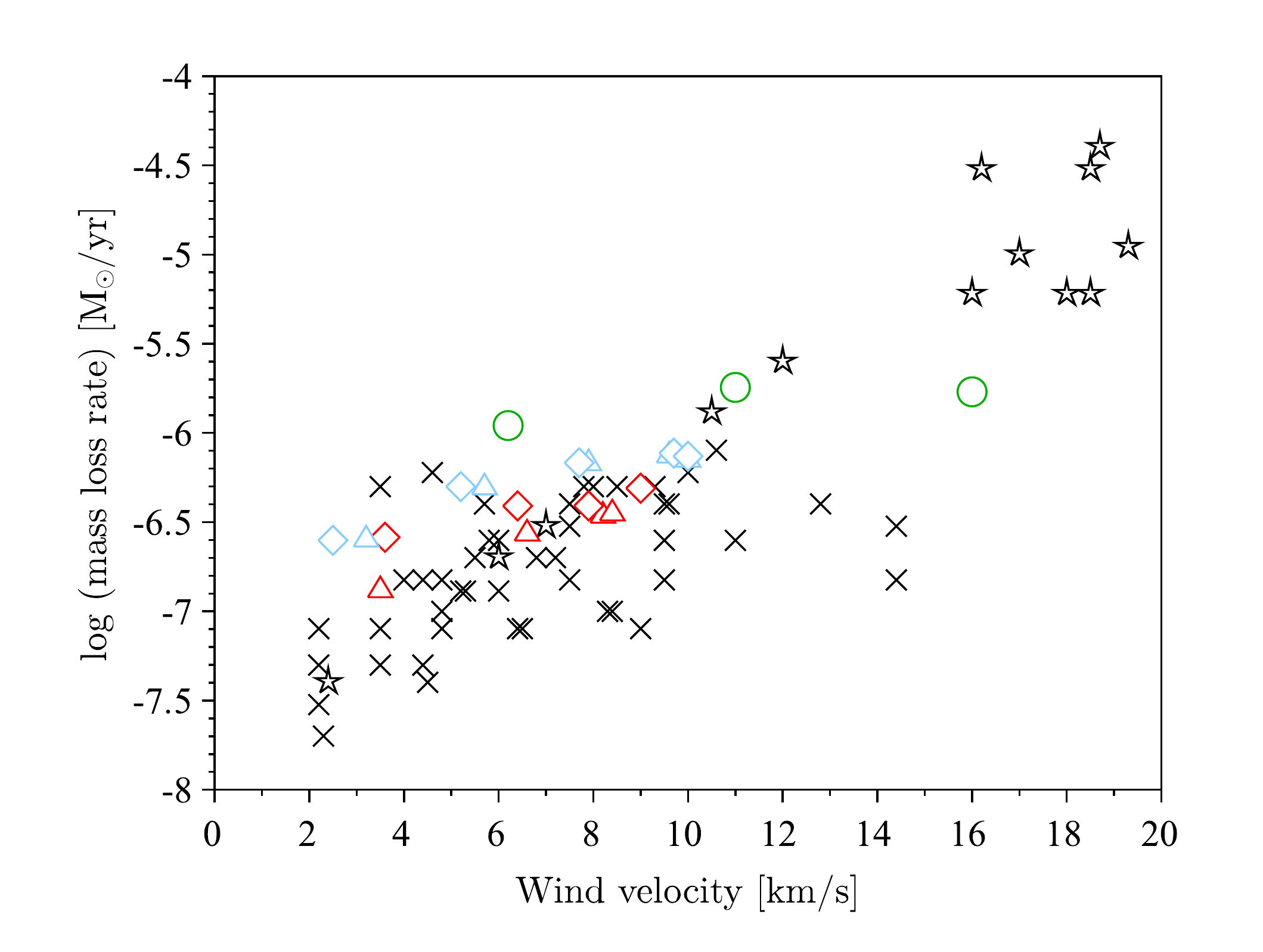}
\includegraphics[width=\hsize]{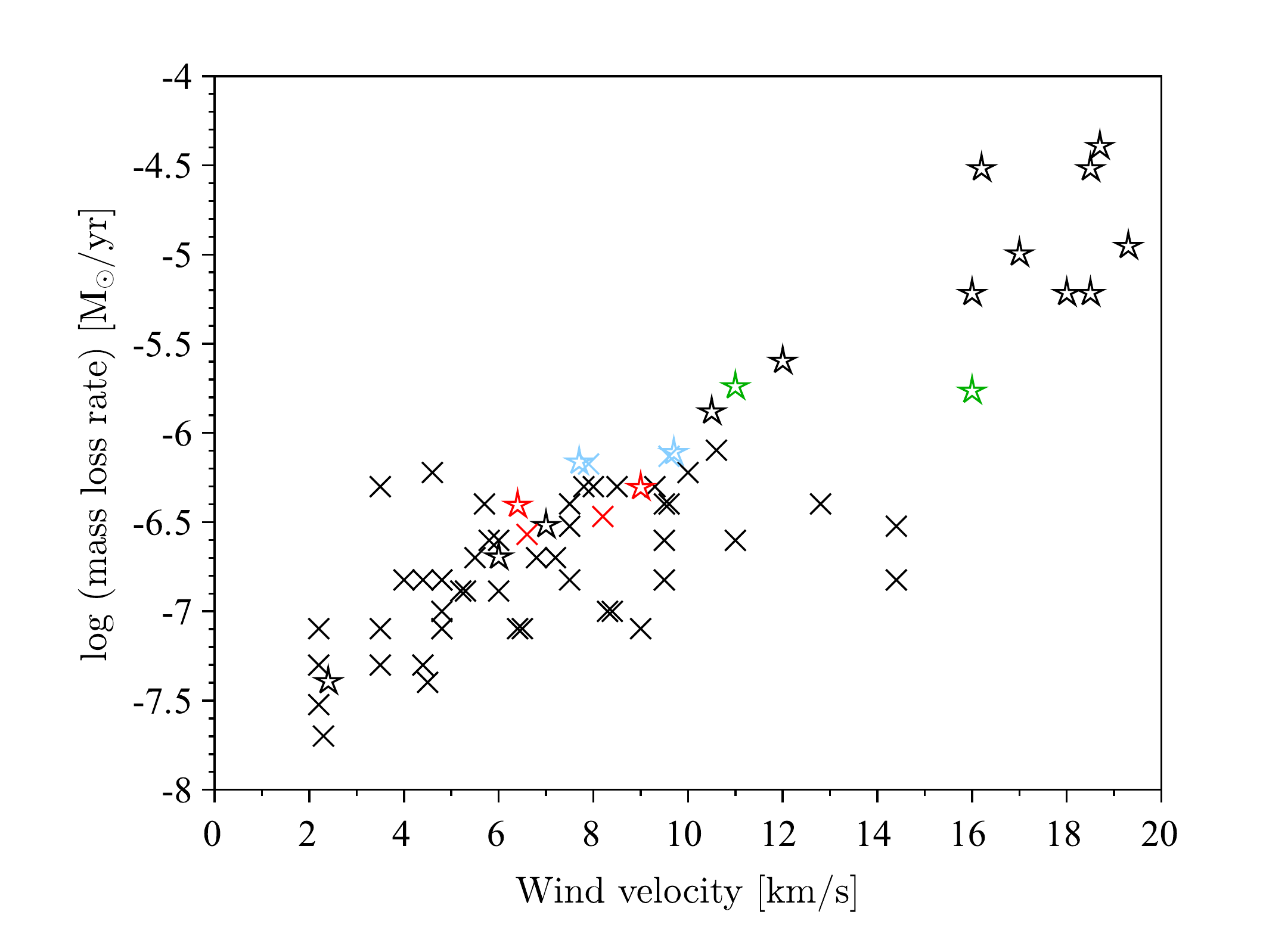}
     \caption{Mass-loss rate versus wind velocity for M-type AGB stars. Observations by \citet{olofetal02} and \citet{gonzetal03} are shown in black (star symbols represent Mira variables, and crosses represent semi-regulars, SRa and SRb); colored symbols represent DARWIN models.   
     {\em Top panel:} All models in Table\,\ref{t_mod} shown, with symbols indicating input parameters. Red symbols mark models of series A and blue symbols models of series B (triangles and diamonds indicate pulsation amplitudes of 3 and 4 km/s, respectively); models of series M2 are shown as green circles.  
     {\em Bottom panel:} Selected models, discussed in Sect.\,\ref{s_obs} (see also Table\,\ref{t_mod_sel}). Symbol shapes indicate variability type (same as observations); red symbols represent series A, blue symbols series B, and green symbols series M2.
             }
      \label{f_dMdt_v_oli}
\end{figure}

\subsection{Trends with model parameters}\label{s_trends}

The conclusions drawn from the detailed analysis of a selected model in the previous section also hold for the other models presented in Table~\ref{t_mod}. They span a considerable range in mass-loss rates and wind velocities, due to differences in the stellar parameters and pulsation properties, as well as seed particle abundances. Nevertheless, the values of Fe/Mg are low for all models (no more than a few percent), due to the self-regulation mechanism via the grain temperature described in Sect.~\ref{s_femg}.

The wind velocities listed in Table~\ref{t_mod} are typically 1-2 km/s higher than in the corresponding models with Fe-free dust (see Appendix\,\ref{app_tab_noFe}, Table~\ref{t_mod_noFe}). The models with the lowest wind velocities for a given set of stellar parameters show the strongest relative increase in radiative acceleration due to additional absorption in the near-IR. The mass-loss rates are basically unchanged compared to models with Fe-free silicates (within the uncertainties of computing temporal means in strongly variable models). This, again, illustrates the point that Fe enrichment is a secondary process, which occurs mostly after the outflow has been initiated by large Fe-free silicate grains. 

Regarding trends with $n_d/n_{\rm H}$ for fixed stellar parameters and pulsation properties, we note that the Fe/Mg values of the grains show an increase with the seed particle abundance. At first, this trend may seem surprising, since larger $n_d/n_{\rm H}$ values mean more grains, and therefore a stronger competition for Fe between grains when forming Fe-enriched mantles. However, it has to be taken into consideration that grain sizes $a_{gr}$ are smaller for higher seed particle abundance (see Table~\ref{t_mod}), and smaller grains have a larger surface/volume ratio. Since Fe enrichment is primarily a surface phenomenon on almost fully grown grains, with the bulk of the grain consisting of Mg$_2$SiO$_4$, there should be a correlation of the Fe/Mg ratio of the grains with their surface/volume ratio. Figure~\ref{f_FeMg_s2v} shows that this is the case: For each series of models that differ by $n_d/n_{\rm H}$ only (same color and symbol shape), there is a clear positive correlation between these two quantities. The only obvious exceptions are models of series A and B with very low wind velocities, about 3-4 km/s (forming large grains with surface/volume ratios below about 8). In these models it is difficult to make a clear-cut distinction between outer atmosphere and inner wind region, and efficient grain growth may stretch over more than one pulsation cycle. 

While Fe/Mg characterizes the composition of individual dust grains, the condensation degrees $f_{\rm Si}$, $f_{\rm Mg}$ and $f_{\rm Fe}$, listed in Table~\ref{t_mod}, represent the total amount of Si, Mg and Fe condensed into dust, and the corresponding depletion of these elements in the gas phase. For a fixed combination of stellar parameters and pulsation amplitude, all three degrees of condensation are higher for higher seed particle abundances $n_d/n_{\rm H}$. Relatively speaking, however, the effect on $f_{\rm Fe}$ is stronger, reflecting the trend in Fe/Mg discussed above. 

\begin{figure*}[ht]
\centering
\includegraphics[width=\hsize]{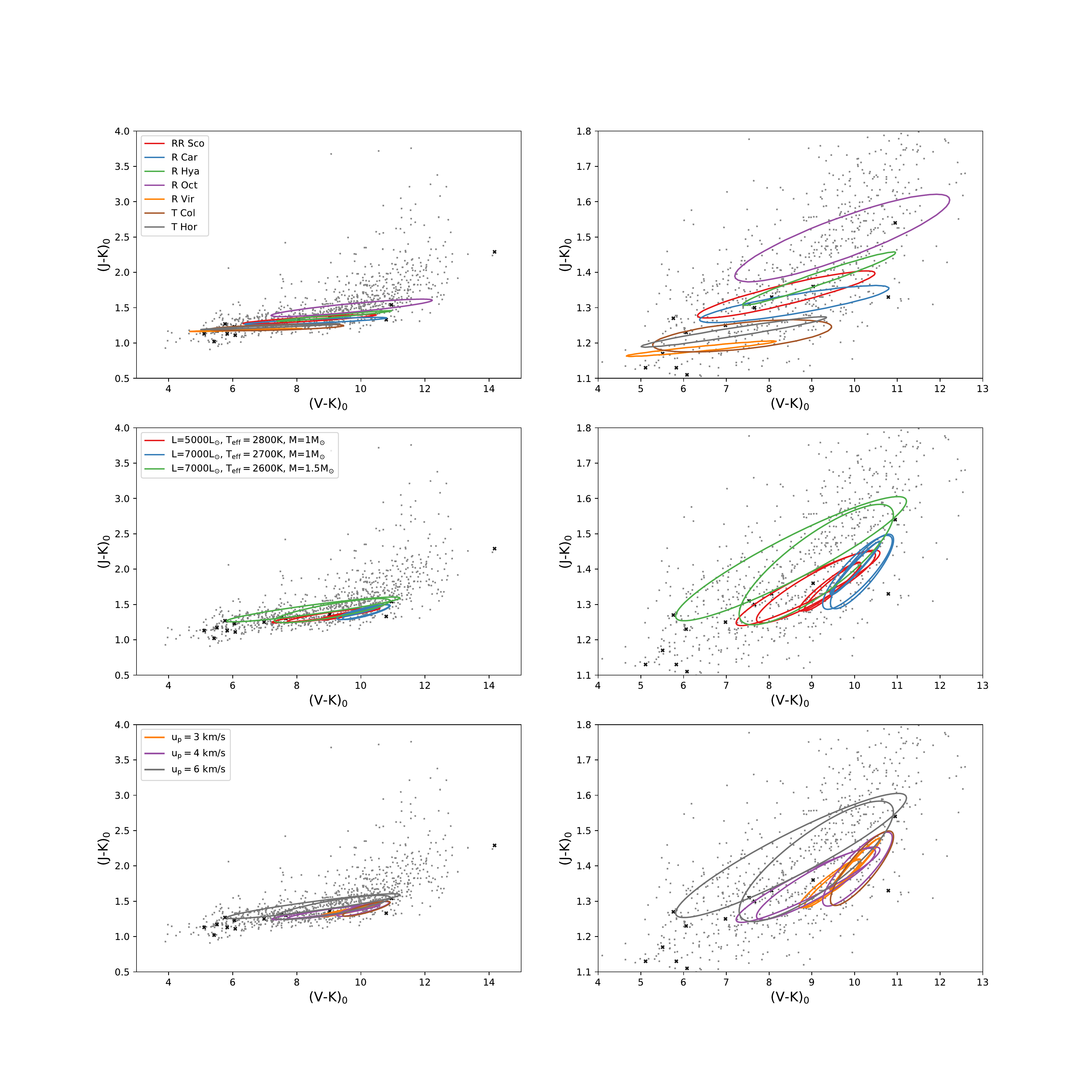}
      \caption{Observed and synthetic visual and near-IR colors of M-type AGB stars. 
      {\em Top panels:} Photometric variations for a sample of observed targets, derived from sine fits of light curves; near-IR data from \citet[][]{whitetal00} have been combined with visual data from \citet[][]{egge75} and \citet[][]{mend67}; see \citet{bladetal13} for details. 
      {\em Middle panels:} Photometric variations in the DARWIN models listed in Table~\ref{t_mod}, represented by loops, with colors calculated from sine fits of the light curves in the same way as for the observational data in the top panels. The lines are color-coded according to stellar parameters. 
      {\em Bottom panels:} Same model results, but color-coded according to pulsation amplitude. 
In all cases, the right panels show the same content as the left panels, zoomed in and centered on the color loops. The single-epoch photometric data shown in the background of all panels represent Galactic bulge Miras \citep[][gray dots]{groeblom05} and field M-type long-period variables \citep[][black crosses]{mend67}.                    }
      \label{f_loops}
\end{figure*}

\begin{figure}
\centering
\includegraphics[width=\hsize]{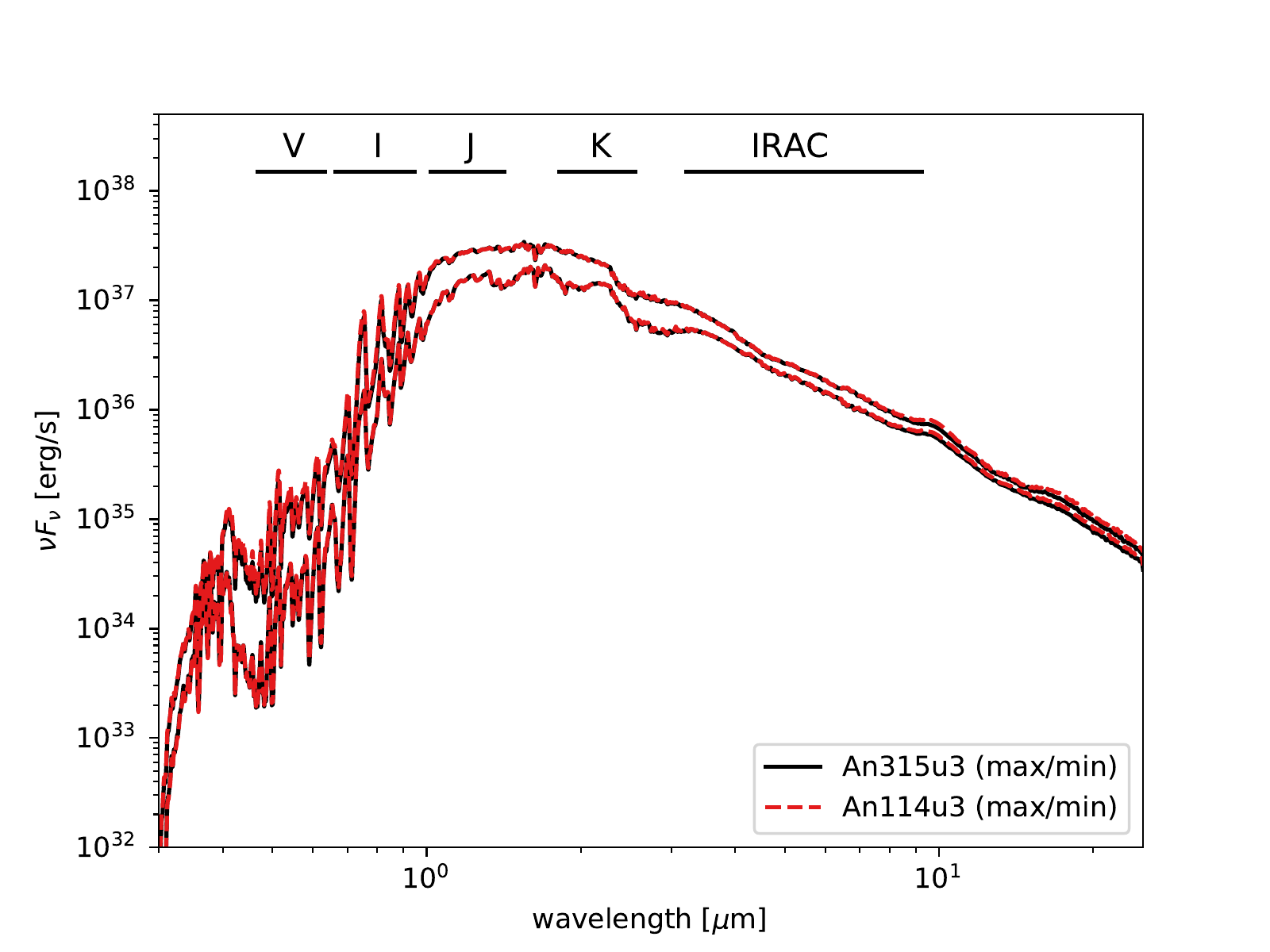}
\includegraphics[width=\hsize]{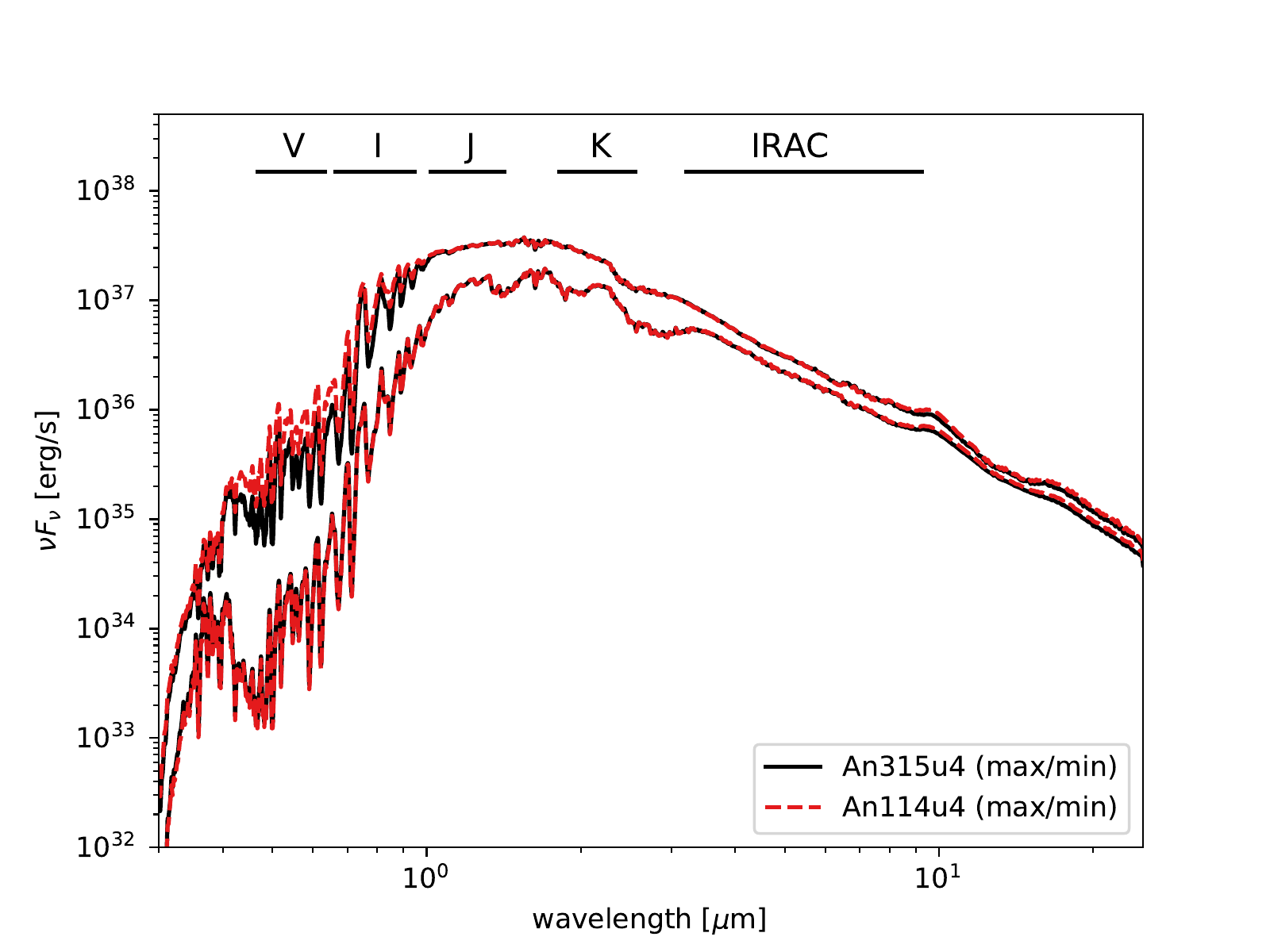}
      \caption{Spectral energy distributions of selected models from series A (see Table\,\ref{t_mod}). Each panel shows SEDs at maximum and minimum light for two models that differ by seed particle abundance only (indicated by line style and color) to illustrate the effect of this parameter. The upper and lower panels, on the other hand, present models with different pulsation amplitudes but otherwise identical parameters. As expected, the models with higher pulsation amplitude (lower panel) show larger spectral variations between maximum and minimum.   
      }
      \label{f_spec_A}
\end{figure}

\subsection{Observable properties}\label{s_obs}

\begin{figure}
\centering
\includegraphics[width=\hsize]{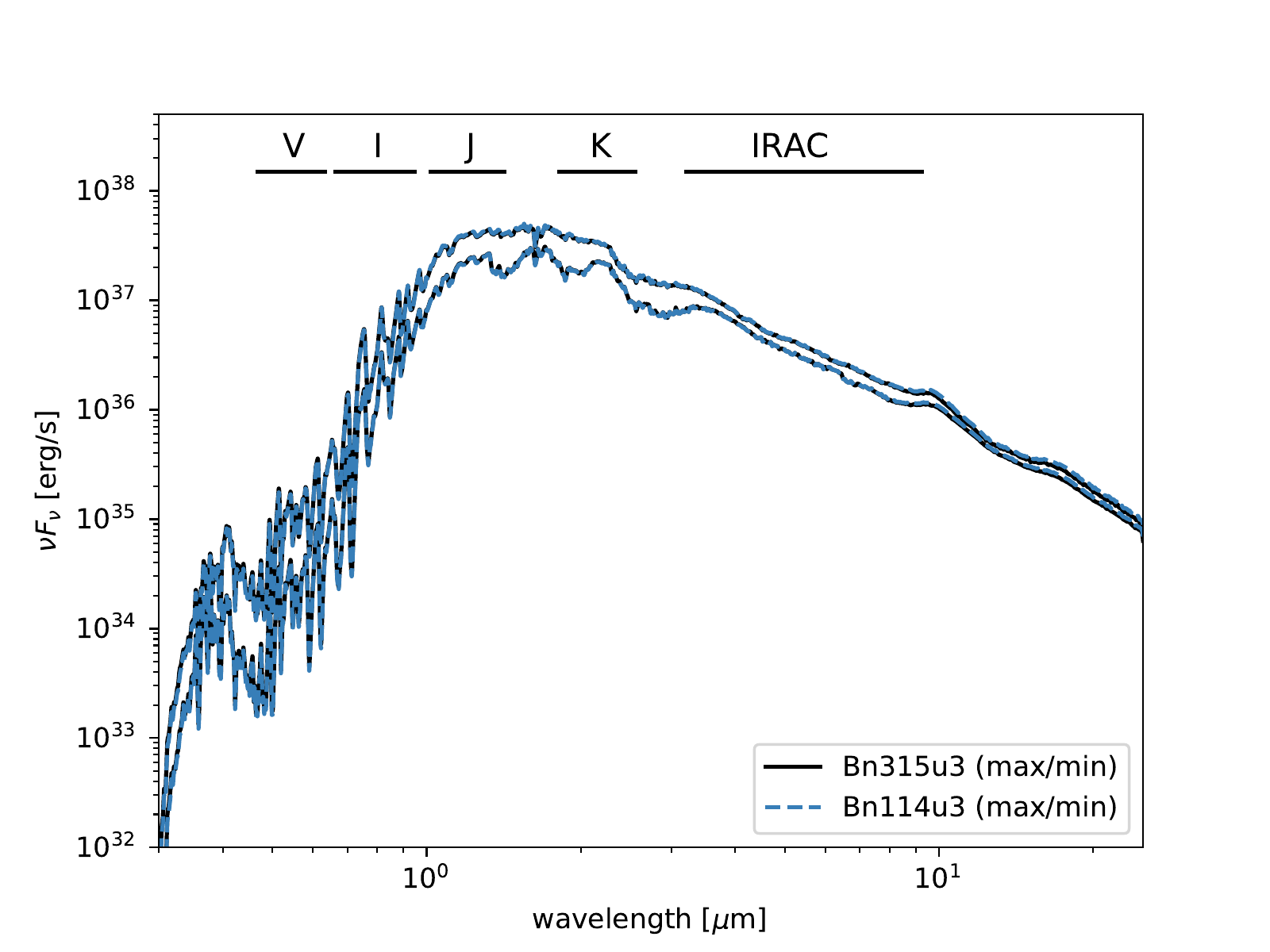}
\includegraphics[width=\hsize]{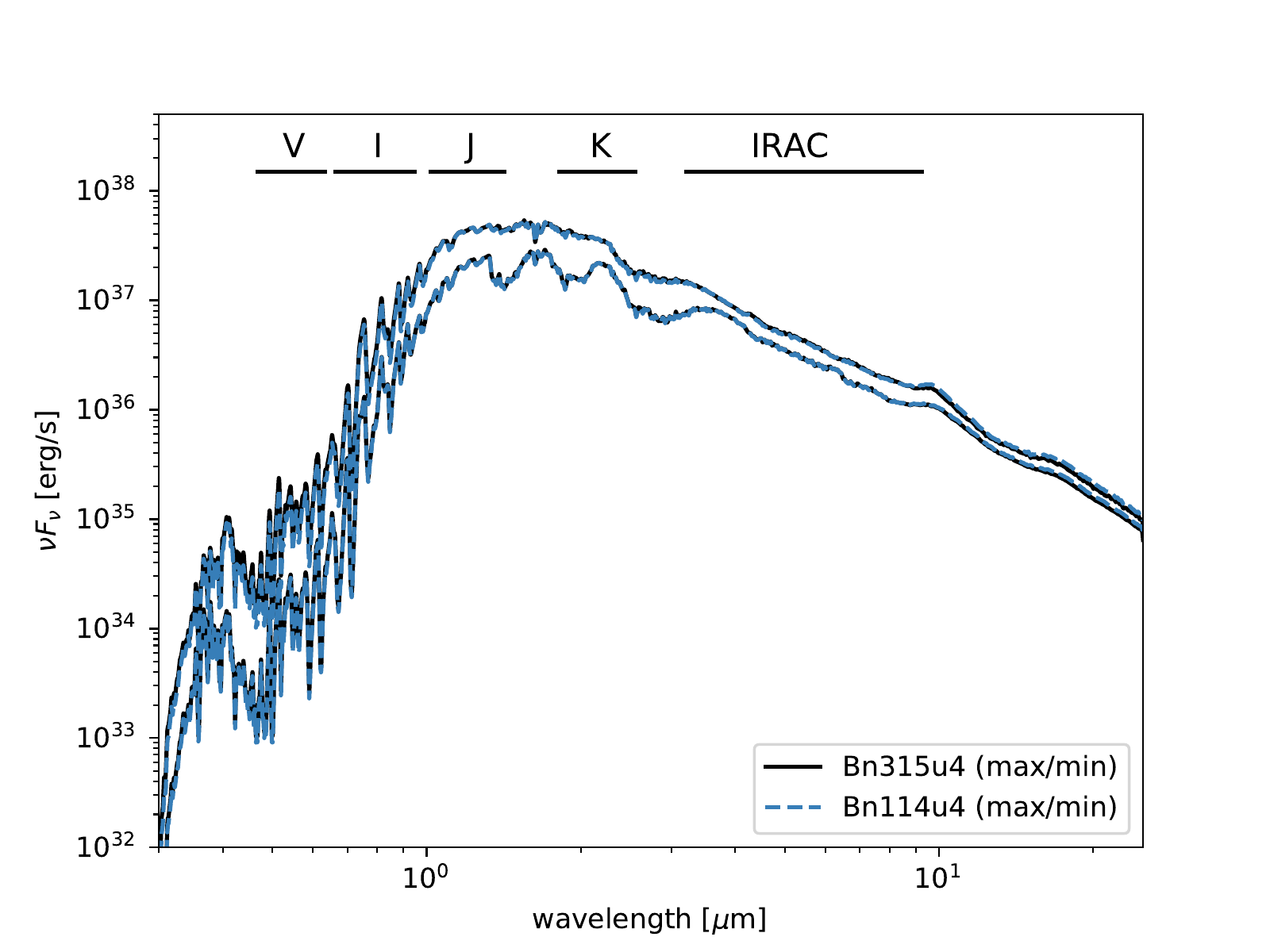}
      \caption{Same as Fig.~\ref{f_spec_A}, but for selected models of series B (see Table\,\ref{t_mod}). 
      }
      \label{f_spec_B}
\end{figure}

\begin{figure}
\centering
\includegraphics[width=\hsize]{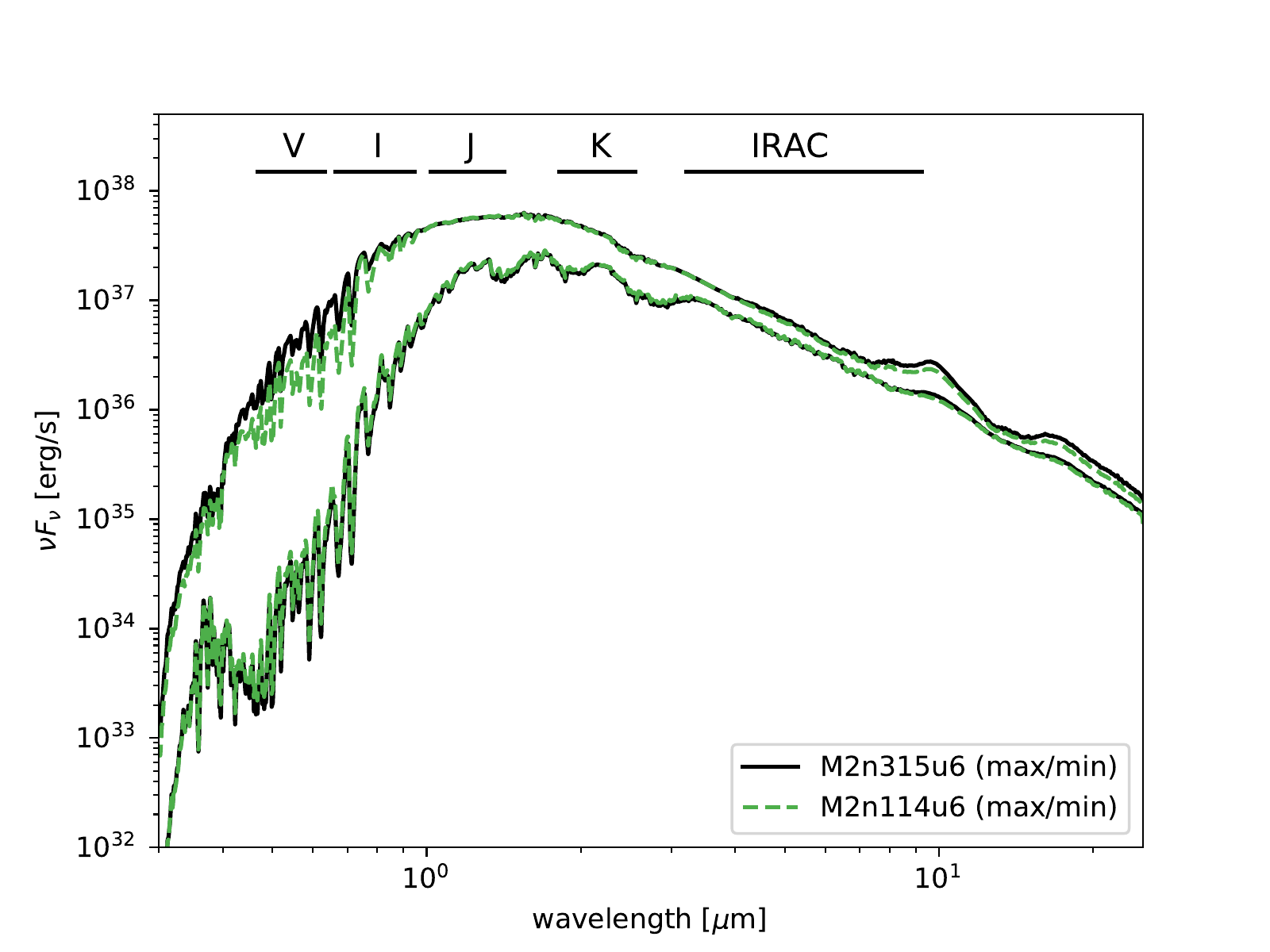}
      \caption{
      Spectral energy distributions at maximum and minimum light for two models of series M2 (Table\,\ref{t_mod}) that differ by seed particle abundance only (indicated by line style and color). 
      }
      \label{f_spec_M2}
\end{figure}

\begin{figure}
\centering
\includegraphics[width=\hsize]{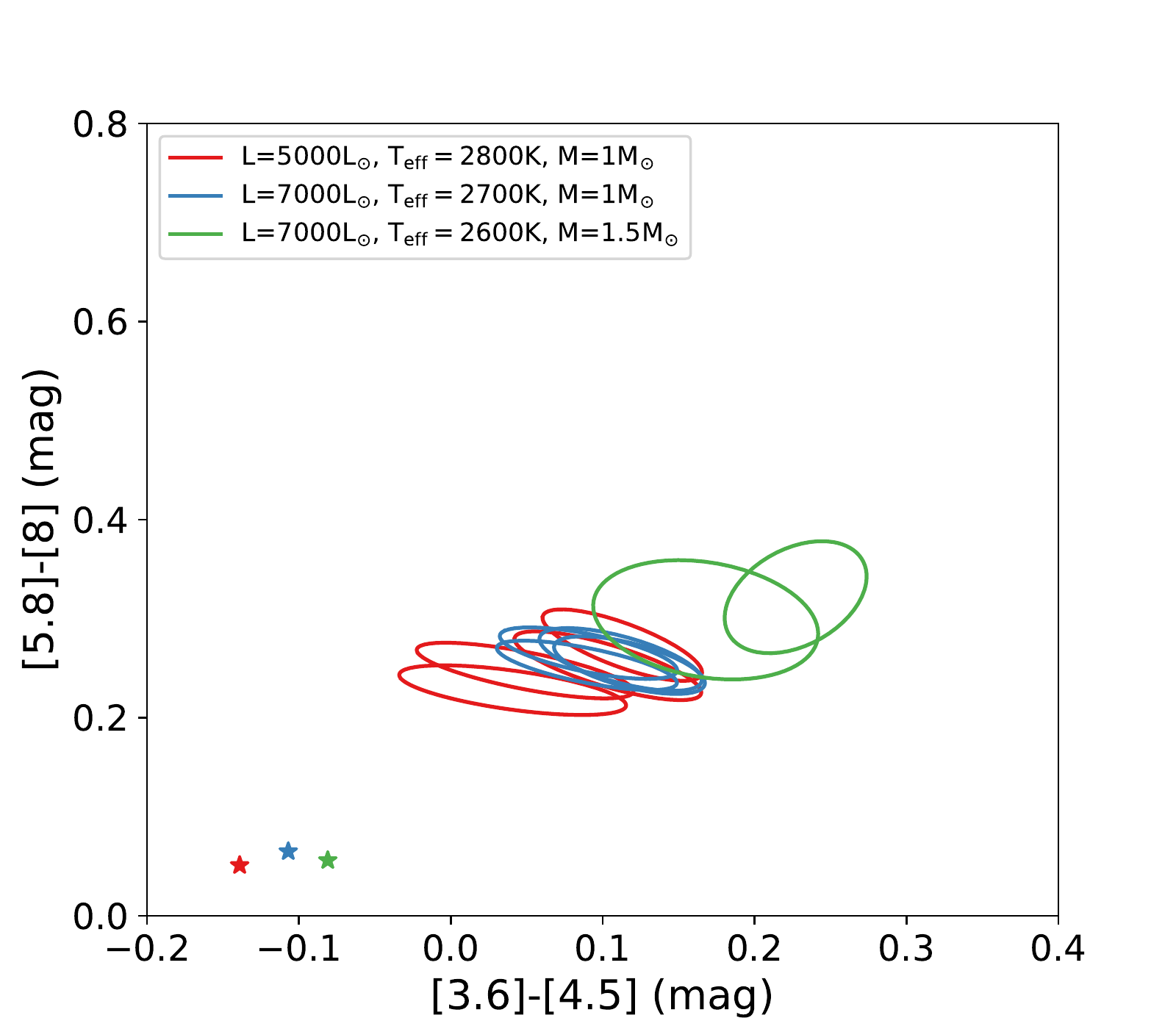}
      \caption{Two-color diagram showing synthetic Spitzer photometry for selected models from Table\,\ref{t_mod}. The loops represent sine fits to the synthetic light curves (as in Fig.~\ref{f_loops} for visual and near-IR colors), while the star-shaped symbols indicate the colors of the corresponding hydrostatic initial models: A, B, and M2.                  }
      \label{f_spitzer}
\end{figure}

%
\begin{table}
\caption{\label{t_mod_sel} 
Subset of the models listed in Table\,\ref{t_mod}, used to illustrate synthetic spectra, photometric properties, and their variability during the pulsation cycle (see Sect.\,\ref{s_obs} and related figures).}
\centering
\begin{tabular}{l|ccccc}
\hline\hline
  & & &  \\
  model & $\Delta M_{\rm bol}$ & $\dot{M}$  & $u_{\rm ext}$ & $\Delta V$ \\
  name & [mag] & [$\Msun$/yr] & [km/s] & [mag] \\
  & & &  \\
\hline
  & & &  \\
  An315u3 / oli & 0.54 & $3 \cdot 10^{-7}$ & 7 & 1.9 & SR \\ 
  An114u3 / oli & 0.54 & $3 \cdot 10^{-7}$ & 8 & 2.0 & SR \\ 
  An315u4 / oli & 0.73 & $4 \cdot 10^{-7}$ & 6 & 3.8 & M \\ 
  An114u4 / oli & 0.73 & $5 \cdot 10^{-7}$ & 9 & 4.3 & M \\ 
  & & &  \\
  Bn315u3 / oli & 0.53 & $7 \cdot 10^{-7}$ & 8 & 1.7 & SR \\ 
  Bn114u3 / oli & 0.53 & $7 \cdot 10^{-7}$ & 10 & 1.7 & SR \\ 
  Bn315u4 / oli & 0.71 & $7 \cdot 10^{-7}$ & 8 & 2.6 & M \\ 
  Bn114u4 / oli & 0.71 & $8 \cdot 10^{-7}$ & 10 & 2.5 & M \\ 
  & & &  \\
  M2n315u6 / oli & 0.95 & $2 \cdot 10^{-6}$ & 11 & 6.2 & M \\ 
  M2n114u6 / oli & 0.95 & $2 \cdot 10^{-6}$ & 16 & 5.6 & M \\ 
  & & &  \\
\hline
\end{tabular}
\tablefoot{Listed here are the bolometric amplitude $\Delta M_{\rm bol}$ (defined by input parameters), the resulting temporal means of the mass-loss rate $\dot{M}$ and the wind velocity $u_{\rm ext}$, and the visual amplitude $\Delta V$ (derived a posteriori with COMA). According to definition, models with $\Delta V > 2.5$ correspond to Mira variables, and smaller amplitudes correspond to semi-regulars (SRa or SRb). 
The model names are constructed in the following way: A, B, or M2 represents a combination of stellar parameters (see Sect.~\ref{s_results}), the letter n followed by a 3-digit number stands for the seed particle abundance (n315 for $n_d/n_{\rm H} = 3 \cdot 10^{-15}$, n114 for $n_d/n_{\rm H} = 1 \cdot 10^{-14}$) and the letter u followed by a number (3, 4, 6) represents the velocity amplitude at the inner boundary (km/s).}
\end{table}

As a first test we confront the basic wind properties of the new models with values derived from observations, in a diagram showing mass-loss rate versus wind velocity (Fig.~\ref{f_dMdt_v_oli}). Due to slightly higher radiative pressure caused by additional absorption in the near-IR, the new models of series A and B with Fe-bearing silicate grains tend to have higher wind velocities than their Fe-free counterparts in Paper I, bringing them into even better agreement with the bulk of observed stars. 
The new models of series M2 also give realistic combinations of wind velocity and mass-loss rate. We note that for all three series the models with intermediate to high values of the seed particle abundance show the best agreement with the observed wind properties (see Fig.~\ref{f_dMdt_v_oli}, bottom panel, and Table\,\ref{t_mod_sel}). In the discussion of spectra, photometry, and variability below, we therefore concentrate on these models.  

In earlier studies we found that synthetic $(J-K)$ and $(V-K)$ colors of models with Mg$_2$SiO$_4$ grains show good agreement with values derived from observational data \citep{bladetal13,bladetal15,hoefetal16}. In particular, the time-dependent behavior of these models during a pulsation cycle was found to be similar to a sample of well-observed Mira variables, that is to say, flat loops in a $(J-K)$ versus $(V-K)$ diagram, with small variations in $(J-K)$ and large variations in $(V-K)$ due to changes in molecular features (mostly TiO and H$_2$O). We interpreted this as a strong indication of wind-driving grains with low absorption efficiency in the visual and near-IR region. Since Fe-bearing silicate grains may cause more absorption than Mg$_2$SiO$_4$ (see Fig.~\ref{f_FeMg_qg}), we need to check how the new models compare to observations at visual and near-IR wavelengths. 

Figure~\ref{f_loops} demonstrates that loops in the ($J-K$) versus ($V-K$) diagram, traced out during a pulsation cycle by the subset of models listed in Table\,\ref{t_mod_sel}, fit well with observed photometric values. As expected, both colors are somewhat redder than for models with pure Mg$_2$SiO$_4$ grains \citep[see Fig.\,\ref{f_loops_noFe} and][Fig.\,4, for comparison]{hoefetal16}, due to the higher absorption of Fe-bearing grains in the visual and near-IR region. 

It should be mentioned here that our experimental models based on composite grains with a Mg$_2$SiO$_4$ mantle growing on top of an Al$_2$O$_3$ core, presented in Paper I, tended to showed flatter loops in the $(J-K)$ versus $(V-K)$ diagram (stronger variation in V) than corresponding models with pure Mg$_2$SiO$_4$ grains. This difference is caused by changes in the transition region between the pulsating atmosphere and the wind, due to grains reaching the critical size regime for wind acceleration faster. In the present paper, we have decided to use silicate dust, only, to isolate the effects of Fe enrichment on the observable properties. Furthermore, adding Al$_2$O$_3$ would entail some technical difficulties (e.g., concerning the computation of optical properties) that are beyond the scope of the current paper, and we intend to return to this point in a future publication. Based on our earlier results, however, we expect that composite grains, with an Fe-enriched silicate mantle growing on top of an Al$_2$O$_3$ core, could lead to larger variations in $(V-K)$. 

Figures~\ref{f_spec_A}, \ref{f_spec_B}, and \ref{f_spec_M2} show spectral energy distributions (SEDs) of the models in Table\,\ref{t_mod_sel}, ranging from visual to mid-IR wavelengths. As for models with Fe-free silicates, the visual and near-IR spectra are dominated by molecular features, notably TiO and H$_2$O. In contrast to these earlier generations of wind models, however, the new models show clear silicate features at 10 and 18 microns (see discussion in Appendix\,\ref{app_obs} and Fig.\,\ref{f_SEDs_noFe}). It is worth noting here that the amount of condensed material is comparable to the corresponding models with Fe-free dust. The addition of a few percent of Fe-bearing silicates mainly affects the dust temperature, leading to grains that are about 200 - 400 K warmer than their Fe-free counterparts (see Figs.\,\ref{f_struct} and \ref{f_struct2}). This is sufficient to produce the characteristic silicate emission features. 

The models in series M2 have the largest pulsation amplitudes, resulting in the strongest variations in $(J-K)$ and $(V-K)$ colors (Fig.~\ref{f_loops}) and the largest $\Delta V$ (Table\,\ref{t_mod_sel}). Figure \ref{f_spec_M2} demonstrates that the variation in  $(V-K)$ is mainly caused by large changes in molecular features during the pulsation cycle. At maximum light both TiO and H$_2$O show much weaker absorption in the visual and near-IR. The considerably higher flux levels at short wavelengths (bluer colors) lead to efficient heating of the dust grains and very pronounced silicate features in the mid-IR. At minimum light, the SED is much redder, and the dust features much weaker. 

Compared to series M2, the models in series A and B have lower pulsation amplitudes, resulting in smaller variations in spectra and colors during the pulsation cycle. Looking at models with $\Delta u_{\rm P} = 4$\,km/s (Figs.\,\ref{f_spec_A} and \ref{f_spec_B}, lower panels), the warmer A-series models show stronger variations in the visual than the B-series models, mostly due to changes in TiO features. The cooler B-series models have strong TiO absorption throughout the pulsation cycle, in contrast to the A-series models, where TiO may get significantly weaker around maximum light, when the atmosphere is warmest. In summary, we see that both the pulsation amplitude and the stellar parameters (in particular effective temperature) affect the variability in the V-band, and the resulting variations in $(J-K)$ and $(V-K)$ colors (Fig.~\ref{f_loops}). 

It should be noted here that $\Delta V$ is a resulting property of the models that depends on the complex interplay of dynamics, luminosity variations, and their effects on molecules and dust. In contrast to $\Delta M_{\rm bol}$, which is set by input parameters, $\Delta V$ has to be determined with detailed a posteriori radiative transfer, using the COMA code. Table\,\ref{t_mod_sel} demonstrates that models with similar $\Delta M_{\rm bol}$ can show very different values of $\Delta V$, illustrating that the amplitude of the visual variation is not necessarily a good proxy of the actual pulsation amplitude. This should be kept in mind when separating Miras and semi-regular variables (SRa) by visual amplitude. 

Figure\,\ref{f_dMdt_v_oli} (lower panel) shows a comparison of variability type for the models in Table\,\ref{t_mod_sel} with observed stars. In the mass-loss rate versus wind velocity diagram, the observed Mira variables ($\Delta V >  2.5$) span the entire range of values, but they are preferably found in the high mass-loss rate and high wind velocity region. In contrast, the semi-regular variables tend to cluster at lower mass-loss rates and wind velocities. The M2 models, both classified as variability type M (see Table\,\ref{t_mod_sel}) fall into the region where only Miras are found in the observed sample. Models of series A and B (variability type M for $\Delta u_{\rm P} = 4$\,km/s, SR for $\Delta u_{\rm P} = 3$\,km/s), on the other hand, are found in the region of the diagram populated by both Miras and semi-regulars.  

With synthetic spectra of M-type DARWIN models showing pronounced mid-IR dust silicate features for the first time, it is also interesting to test photometric fluxes and colors at longer wavelengths than in previous papers, based on models with Fe-free silicate dust. Figure \ref{f_spitzer} shows synthetic photometry in {\em Spitzer} filters  for the same selection of models as in Fig.~\ref{f_loops}. The loops represent variations during a pulsation cycle, while the star-shaped symbols indicate the colors of the corresponding hydrostatic initial models A, B, and M2. The variations in the 3.6, 4.5 and 5.8 micron fluxes are mainly caused by changes in molecular features (in particular H$_2$O) during the pulsation cycle, while the 8 micron flux is partly affected by the nearby silicate feature. The dynamic models with dust-driven winds are significantly redder in both colors than their dust-free hydrostatic counterparts, and they fall into a part of the diagram associated with M-type AGB stars 
\citep[e.g.,][]{boyeetal11,sargetal11,joneetal17}


\section{Discussion}\label{s_discussion}

Summarizing the results presented above, we see that the Fe-content of the wind-driving silicate grains is limited by two factors: close to the star by radiative heating due to larger absorption efficiency with higher Fe/Mg; further away, by a critical slowing down of grain growth due to quickly decreasing gas densities in the wind. The resulting Fe/Mg ratios are low, no more than a few percent. Nevertheless, the synthetic spectra show distinctive silicate features around 10 and 20 microns, due to higher grain temperatures, in contrast to earlier DARWIN models with Fe-free dust. 

In this context it is worth noting that we have assumed very favorable conditions for Fe enrichment, that is, a sticking coefficient of $\alpha_{\rm Fe} = 1$ (see Sect.~\ref{s_oli}). A significantly lower value of $\alpha_{\rm Fe}$ would lead to lower Fe/Mg, and consequently to a weakening or even disappearance of the mid-IR silicate features. A point in favor of a high sticking probability is that the bulk of the dust consists of Mg$_2$SiO$_4$, leading to a much lower grain temperature at a given distance than for a material with a high content of Fe.  

Another underlying assumption of the new models is that the internal structure of the grains can be represented by a core of Mg$_2$SiO$_4$ surrounded by a geometrically thin mantle of MgFeSiO$_4$ for the purpose of calculating optical properties (see Sect.~\ref{s_tdust}). This is consistent with the finding that the enrichment of the silicate dust with Fe is a secondary process, taking place in the innermost wind region on the surface of large Mg$_2$SiO$_4$ grains that are initiating the outflow. Figure\,\ref{f_xsurf} shows the estimated composition at the grain surface, expressed by $x_{\rm surf}$, as defined by Eq.\,(\ref{e_xsurf}). The derived values are within the expected range (see Sect.\,\ref{s_tdust} and Appendix\,\ref{app_x05}), confirming that $x=0.5$  (corresponding to MgFeSiO$_4$) is a reasonable approximation for the mantle material. 

The results presented in Sect.~\ref{s_results} are consistent with earlier studies that take the effects of composition-dependent opacities on dust condensation and grain temperatures into account. \cite{woit06} explored silicate formation in an outflow driven by a preset force, concluding that Fe-bearing silicates can only form at distances of several stellar radii, too far from the stellar surface to initiate a wind. \cite{bladetal15} showed that the absence of mid-IR silicate features in the spectra of wind models based on pure Mg$_2$SiO$_4$ grains is not caused by a lack of mid-IR opacity (see also Fig.~\ref{f_FeMg_qg}), but a consequence of low grain temperatures. Setting an artificial lower limit to the falling grain temperature in the outflow when computing synthetic spectra from otherwise unmodified wind models lead to distinctive mid-IR silicate features. Another test presented by \cite{bladetal15} was to add thin mantles of MgFeSiO$_4$ on the original Mg$_2$SiO$_4$ grains during {\em a posteriori} radiative transfer, with a given ratio of mantle to total grain radius, starting at a distance from the star where such mantles could be thermally stable. The cases of 1\% and 5\% mantle/total grain radius are roughly comparable to the Fe/Mg values in the models presented here. 

A consistent treatment of the feedback between composition and radiative heating during the growth of silicate grains leads to a gradual increase in Fe/Mg and a more complex dependence of the dust temperature on distance from the star than in the simple test cases presented by \cite{bladetal15}. As shown in Fig.~\ref{f_struct} the growing Fe/Mg ratio in the innermost wind region is accompanied by dust temperatures that are much higher than for Fe-free silicate grains. The strongest effect on temperature is found around $4-5 \, \Rstar$ (see Fig.~\ref{f_struct}, middle right panel), a region that is remarkably similar to 
silicate condensation distances derived from spectro-interferometric measurements \citep[see][and references therein]{karoetal13}. Our new models lead us to the conclusion that the pronounced mid-IR emission seen in the spatially resolved observations around $4-5 \, \Rstar$ is caused by Fe/Mg rising to a few percent in this region, while the bulk of Fe-free silicate dust is formed around $2 \, \Rstar$, initiating the outflow. Indeed, spectro-interferometric observations of RT Vir by \cite{sacuetal13} show a weak silicate feature at about $2 \, \Rstar$. In Fig.~\ref{f_struct2} we show the time-dependent radial structure of a model with different stellar and pulsation parameters, compared to the one in Fig.\,\ref{f_struct}, resulting in different wind and dust properties. It demonstrates that the strong heating around $4-5 \, \Rstar$ due to Fe enrichment is not specific to the model presented in Fig.~\ref{f_struct}. As pointed out by \cite{bladetal17}, mid-IR spectro-interferometric observations may be a powerful tool for tracing changes in grain composition with distance from the star. A discussion of the corresponding synthetic observables for the new DARWIN models presented here, however, is beyond the scope of this paper and will be the subject of a future publication.  


\section{Summary and conclusions}\label{s_concl}

Magnesium-iron silicates have long been regarded as good candidates for driving the winds of M-type AGB stars, considering the abundances of relevant elements (Si, Mg, Fe, O) and the prominent mid-IR silicate features observed in circumstellar dust shells. As shown in earlier papers, DARWIN models of winds driven by photon scattering on large Mg$_2$SiO$_4$ grains do produce realistic mass-loss rates and wind velocities, as well as visual-to-near-IR spectra that compare well with observations \citep{hoef08,bladetal15,bladetal19,hoefetal16}. 
However, due to low grain temperatures, their synthetic spectra show no mid-IR silicate features. 

Here we have presented new DARWIN models that allow for the growth of silicate grains with a variable Fe/Mg ratio in the dynamical atmosphere and wind. The self-regulating feedback between grain composition and corresponding radiative heating, in combination with quickly falling densities in the stellar wind, leads to low final values of Fe/Mg, typically a few percent. Nevertheless, the new models show distinct silicate features around 10 and 18 microns. The effect of the Fe enrichment on visual and near-IR photometry is moderate, and the new DARWIN models agree well with observations regarding ($J-K$) versus ($V-K$) colors (including variations during a pulsation cycle) and {\em Spitzer} color-color diagrams. 

The gradual Fe enrichment of silicate grains in the inner wind region should produce observable signatures in mid-IR spectro-interferometrical measurements \citep[see][for a discussion]{bladetal17}, providing additional possibilities for testing the scenario suggested by the new DARWIN models. In this context, we point out that the strongest effect of the rising Fe/Mg value on grain temperature is found at around $4-5\,\Rstar$, which is remarkably similar to silicate condensation distances derived from spatially resolved observations, while the bulk of Fe-free silicate dust forms at around $2\,\Rstar$ in the models and drives the wind.  

It is important to note that the enrichment of the silicate dust with Fe is a secondary process, taking place in the stellar wind on the surface of large Fe-free grains that can form closer to the star and initiate the outflow. Therefore, the mass-loss rates are basically unaffected, while the wind velocities tend to be higher by a few km/s than in corresponding models with Fe-free silicate dust. We conclude that mass-loss rates derived from existing DARWIN model grids for M-type AGB stars, based on Fe-free silicate grains \citep{bladetal19}, can safely be used in stellar evolution models since the values are not significantly affected by the inclusion of Fe in the silicate grains, which occurs after the outflow has been established. When comparing synthetic spectra and photometry to observations, however, the effects of Fe enrichment should not be neglected. 


\begin{acknowledgements}
      This work is part of a project that has received 
      funding from the European Research Council (ERC) 
      under the European Union’s Horizon 2020 research and innovation programme 
      (Grant agreement No. 883867, project EXWINGS), 
      the Swedish Research Council ({\it Vetenskapsrådet}, grant number 2019-04059), 
      and the Schönberg donation.   
      The computations of spectra and photometry were enabled by resources provided by the 
      Swedish National Infrastructure for Computing (SNIC) at UPPMAX 
      partially funded by the Swedish Research Council through grant agreement no. 2018-05973. 
\end{acknowledgements}


\bibliographystyle{aa} 
\bibliography{hoefner} 


\begin{appendix} 

\section{Internal structure of dust grains}\label{app_x05}

\begin{figure}
\centering

\includegraphics[width=\hsize]{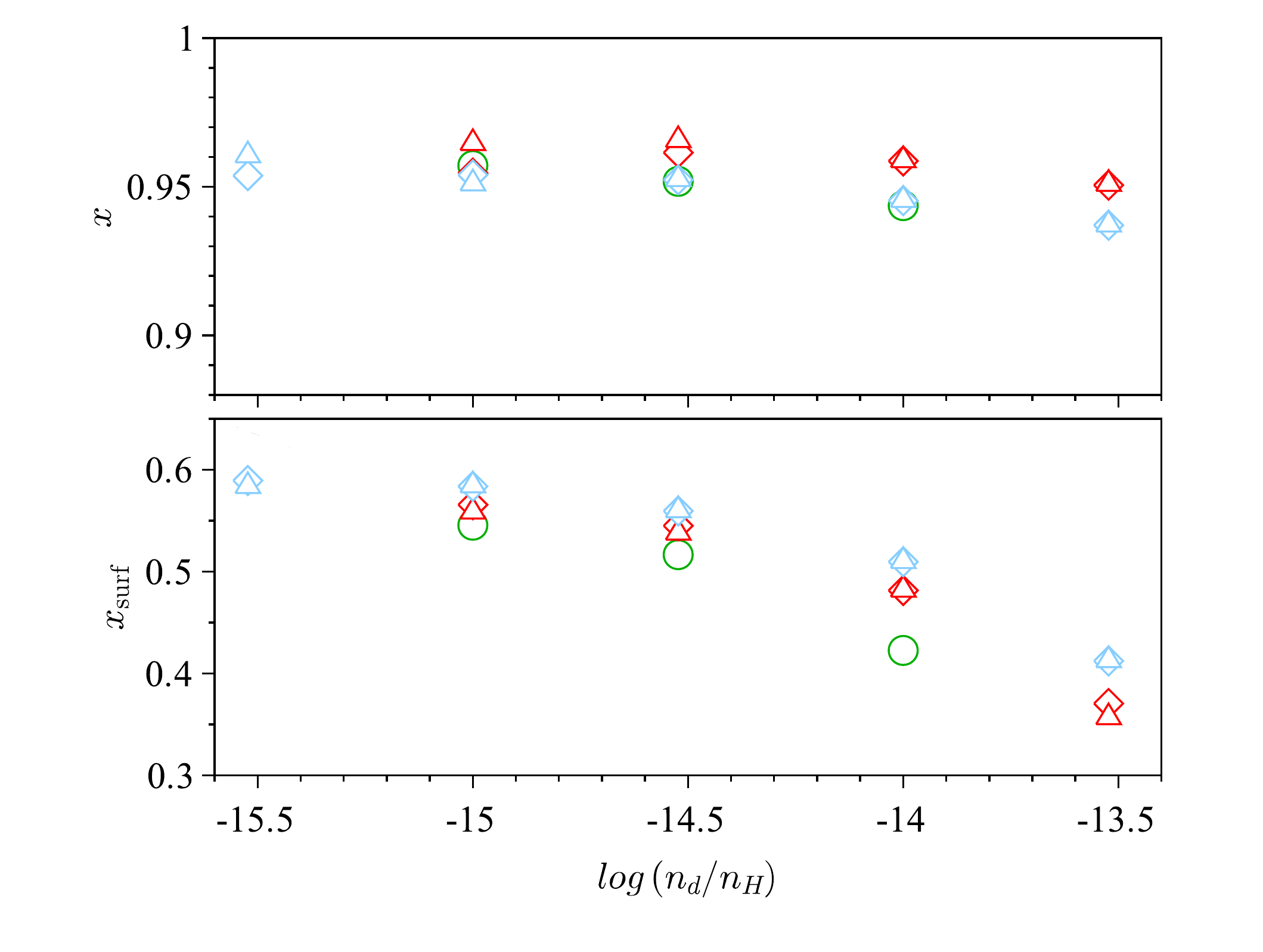}
     \caption{
     Composition of silicate grains in the DARWIN models. 
     {\em Upper panel:} Total value $x = (1 + {\rm Fe/Mg})^{-1}$ for a grain, with Fe/Mg taken from Table~\ref{t_mod}.
     {\em Lower panel:} Estimated value at the grain surface $x_{\rm surf}$ as given by Eq.\,(\ref{e_xsurf}), using condensation degrees $f_{\rm Mg}$ and $f_{\rm Mg}$ from Table~\ref{t_mod} in Eq.\,(\ref{e_FeMgsurf}) for $({\rm Fe/Mg})_{\rm surf}$ (to account for the depletion of Fe and Mg in the gas phase in the growth rates). In both panels, red symbols mark models of series A and blue symbols models of series B (triangles and diamonds indicate pulsation amplitudes of 3 and 4 km/s, respectively); models of series M2 are shown as green circles.
     }
      \label{f_xsurf}
\end{figure}

In this paper, we describe the growth of olivine-type silicate grains with a variable composition, characterized by $x = {\rm Mg / (Mg+Fe)} = (1+{\rm Fe/Mg})^{-1}$. The total value of Fe/Mg for a grain results directly from solving the equations given in Sect.\,\ref{s_oli}, but keeping track of grain growth history, and thereby the detailed interior structure of the silicate grains, in DARWIN models is a nontrivial task, as mentioned in Sect.\,\ref{s_tdust}. When computing the optical properties of the composite grains, we used a simplified core-mantle approach for describing their internal structure, which allowed us to use standard Mie theory. Taking constraints of radiative heating on grain growth into consideration, we assumed a core consisting of Mg$_2$SiO$_4$, surrounded by a mantle of Fe-bearing silicates (with the thickness of the mantle given by Fe/Mg). 

Choosing a value of $x = 0.5$ (equal amounts of Fe and Mg) as an approximation for the mantle material can be motivated by the following consideration: Comparing the growth rates of Mg$_2$SiO$_4$ and Fe$_2$SiO$_4$ (i.e., Eq.\,(\ref{e_jgr_ol_mg}) and Eq.\,(\ref{e_jgr_ol_fe}), respectively) gives an estimate of the surface composition of the growing grains in the region where temperatures have fallen below the condensation thresholds of both species, and decomposition rates can be neglected. The ratio of Fe/Mg in the condensing material at the grain surface is then given by 
\begin{equation}
  ({\rm Fe/Mg})_{\rm surf} = 
        \frac{\alpha_{\rm Fe} \, v_{\rm Fe} \, n_{\rm Fe}}{\alpha_{\rm Mg} \, v_{\rm Mg} \, n_{\rm Mg}} 
      = \frac{\alpha_{\rm Fe} \, v_{\rm Fe} \, \varepsilon_{\rm Fe} \, (1 - f_{\rm Fe})}{\alpha_{\rm Mg} \, v_{\rm Mg} \, \varepsilon_{\rm Mg} \, (1-f_{\rm Mg})} ,
\end{equation}
where we have expressed the gas-phase densities $n_{\rm Fe}$ and $n_{\rm Mg}$ by the respective elemental abundances $\varepsilon_{\rm Fe}$ and $\varepsilon_{\rm Fe}$, 
multiplied by a factor describing depletion by condensation ($f_{\rm Fe}$ and $f_{\rm Mg}$ denote the condensation fractions of Fe and Mg). Expressing the ratio of the velocities $v_{\rm Fe}/v_{\rm Mg}$ as $\sqrt{ m_{\rm Mg}/m_{\rm Fe}}$ and using $\alpha_{\rm Fe} = \alpha_{\rm Mg} = 1$ (see Sect.\,\ref{s_oli}), we obtain
\begin{equation}\label{e_FeMgsurf}
  ({\rm Fe/Mg})_{\rm surf} = 
        \sqrt{ \frac{m_{\rm Mg}}{m_{\rm Fe}} } \, 
        \frac{\varepsilon_{\rm Fe} \, (1-f_{\rm Fe})}{\varepsilon_{\rm Mg} \, (1-f_{\rm Mg})} 
      \approx 0.55 \, \frac{(1-f_{\rm Fe})}{(1-f_{\rm Mg})} 
,\end{equation}
where we have used $m_{\rm Mg}$ = 24, $m_{\rm Fe}$ = 56, $\varepsilon_{\rm Mg} = 3.8 \cdot 10^{-5}$, and $\varepsilon_{\rm Fe} = 3.18 \cdot 10^{-5}$. Defining the corresponding surface value of $x$ as 
\begin{equation}\label{e_xsurf}
  x_{\rm surf} = ( 1 + ({\rm Fe/Mg})_{\rm surf} )^{-1} 
\end{equation}
provides a rough idea about the composition of the Fe-bearing mantle of the silicate grains. 
Using typical values of $f_{\rm Mg}$ in the range of $0.3 - 0.7$ (see Table\,\ref{t_mod_noFe}, listing results from Paper I), and expected values of $f_{\rm Fe}$ around a few percent \citep{bladetal15,bladetal17}, we estimate values of (Fe/Mg)$_{\rm surf}$ to be in the range of $1.8 - 0.8$. This translates into approximate values of $0.3 - 0.6$  for $x_{\rm surf}$, bracketing the assumed mantle value of $0.5$.

Figure\,\ref{f_xsurf} (lower panel) shows the values of $x_{\rm surf}$ for the new models presented in this paper. The condensation degrees $f_{\rm Fe}$ and $f_{\rm Mg}$ are taken from Table~\ref{t_mod}, accounting for the depletion of Fe and Mg in the gas in the growth rates in Eq.\,(\ref{e_FeMgsurf}). The values of $x_{\rm surf}$ in Fig.\,\ref{f_xsurf} agree well with the expected range of values based on previous studies, as estimated above. The value of $x=0.5$ assumed for the mantle material when calculating optical properties of the grains is confirmed to be a reasonable approximation. 

\section{Models with Fe-free silicates}\label{app_tab_noFe}

\begin{table}
\caption{\label{t_mod_noFe}
Properties of DARWIN models with outflows driven by photon scattering on Fe-free silicate grains. 
}
\centering
\begin{tabular}{l|ccccc}
\hline\hline
  & & & \\
  model &  $\dot{M}$  & $u_{\rm ext}$  & $f_{\rm Si}$  & $f_{\rm Mg}$ & $a_{gr}$ \\
  name  & [$\Msun$/yr] & [km/s] &  &  & [$\mu$m] \\ 
  & & & \\
\hline
  & & & \\
  An115u3 & $1 \cdot 10^{-7}$ & 2 & 0.17 & 0.32 & 0.47 \\
  An315u3 & $3 \cdot 10^{-7}$ & 5 & 0.20 & 0.37 & 0.34 \\
  An114u3 & $4 \cdot 10^{-7}$ & 7 & 0.27 & 0.50 & 0.25 \\
  An314u3 & $4 \cdot 10^{-7}$ & 7 & 0.38 & 0.71 & 0.20 \\
  & & & \\
  An115u4 & $3 \cdot 10^{-7}$ & 3 & 0.17 & 0.32 & 0.47 \\
  An315u4 & $5 \cdot 10^{-7}$ & 6 & 0.20 & 0.37 & 0.34 \\
  An114u4 & $6 \cdot 10^{-7}$ & 9 & 0.27 & 0.50 & 0.25 \\
  An314u4 & $5 \cdot 10^{-7}$ & 8 & 0.37 & 0.69 & 0.20 \\
  & & & \\
\hline
  & & & \\
  Bn316u3 & $3 \cdot 10^{-7}$ & 2 & 0.13 & 0.24 & 0.64 \\
  Bn115u3 & $5 \cdot 10^{-7}$ & 5 & 0.14 & 0.26 & 0.44 \\
  Bn315u3 & $7 \cdot 10^{-7}$ & 7 & 0.17 & 0.32 & 0.33 \\
  Bn114u3 & $8 \cdot 10^{-7}$ & 8 & 0.24 & 0.45 & 0.24 \\
  Bn314u3 & $8 \cdot 10^{-7}$ & 8 & 0.35 & 0.65 & 0.19 \\
  & & & \\
  Bn316u4 & $3 \cdot 10^{-7}$ & 2 & 0.13 & 0.24 & 0.64 \\
  Bn115u4 & $5 \cdot 10^{-7}$ & 4 & 0.13 & 0.24 & 0.43 \\
  Bn315u4 & $7 \cdot 10^{-7}$ & 6 & 0.17 & 0.32 & 0.33 \\
  Bn114u4 & $8 \cdot 10^{-7}$ & 8 & 0.24 & 0.45 & 0.24 \\
  Bn314u4 & $8 \cdot 10^{-7}$ & 8 & 0.34 & 0.64 & 0.19 \\
  & & & \\
\hline
  & & & \\
  M2n115u6 & $1 \cdot 10^{-6}$ & 5 & 0.19 & 0.36 & 0.49 \\
  M2n315u6 & $2 \cdot 10^{-6}$ & 12 & 0.23 & 0.43 & 0.36 \\
  M2n114u6 & $2 \cdot 10^{-6}$ & 14 & 0.32 & 0.60 & 0.27 \\
  & & & \\
\hline
\end{tabular}
\tablefoot{
The capital letters in the model name denote different combinations of stellar parameters (see Sect.~\ref{s_results}), the letter n followed by a 3-digit number stands for the seed particle abundance (i.e., n316 for $n_d/n_{\rm H} = 3 \cdot 10^{-16}$, n115 for $n_d/n_{\rm H} = 1 \cdot 10^{-15}$, etc.) and the letter u followed by a number represents the velocity amplitude at the inner boundary $\Delta u_{\rm P}$  (km/s). The resulting wind and dust properties listed here are temporal means of the mass-loss rate $\dot{M}$, the wind velocity $u_{\rm ext}$, the fraction of Si condensed into grains $f_{\rm Si}$, the fraction of Mg condensed into grains $f_{\rm Mg}$, and the grain radius $a_{gr}$. The values for models of series A and B listed here are taken from Paper I \citep{hoefetal16}.
}
\end{table}

Wind properties and other results of models with Fe-free silicate dust are listed in Table~\ref{t_mod_noFe}, for comparison with the new DARWIN models presented in Table\,\ref{t_mod}. The values given are time averages of the respective quantities over several hundred pulsation periods, taken at the outer boundary of the models (at about $20-30 \, \Rstar$). The models of series A and B with winds driven by Fe-free silicate grains have earlier been presented in Paper I, while the models of series M2, both with and without Fe enrichment have been produced for this paper. 

The mass-loss rates are similar for both types of models, since the wind is initiated by Fe-free grains. The wind velocities, on the other hand tend to be about 1-2 km/s higher in the new models with gradual Fe enrichment in the wind, due to higher absorption at visual and near-IR wavelengths (see Fig.~\ref{f_FeMg_qg}). Grain sizes are very similar for corresponding models with and without Fe enrichment, since a few percent of Fe-rich silicates added to the grain volume correspond to an even smaller increase in radius.

\section{Fe enrichment and observables}\label{app_obs}

\begin{figure}
\centering
\includegraphics[width=\hsize]{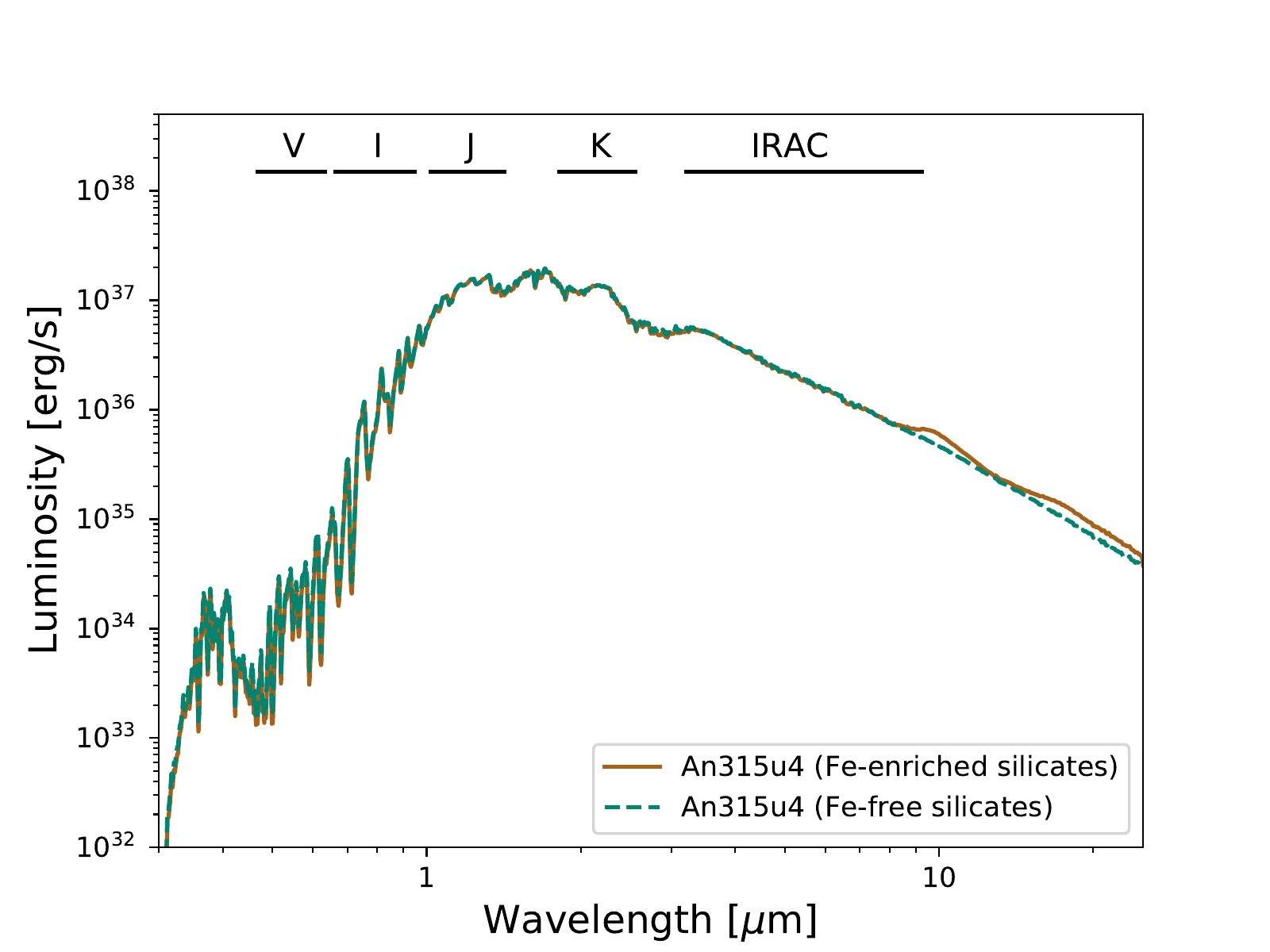}
     \caption{Spectral energy distribution at minimum light of model An315u4 and the corresponding model with Fe-free silicate dust (see text). While containing a similar amount of silicate dust (resulting in comparable  mid-IR opacities), the latter model, due to low grain temperatures, does not show the characteristic silicate features around 10 and 18 micron.
     }
      \label{f_SEDs_noFe}
\end{figure}

\begin{figure}
\centering
\includegraphics[width=\hsize]{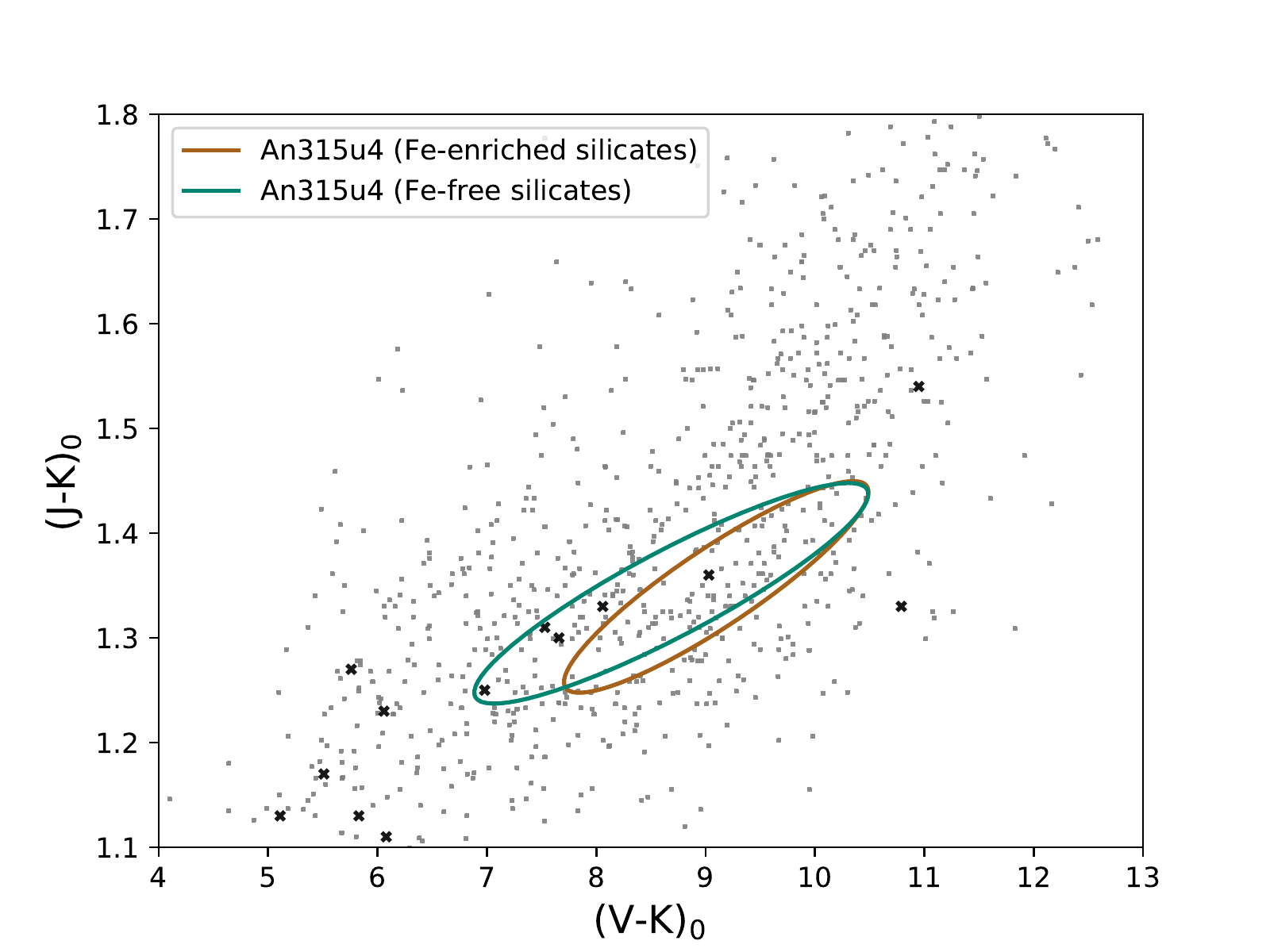}
     \caption{Photometric variations in model An315u4 and the corresponding model with Fe-free silicate dust (see text). The latter model shows bluer $(V - K)$ colors due to lower absorption in the visual region. The single-epoch photometric data shown in the background represent Galactic bulge Miras \citep[][gray dots]{groeblom05} and field M-type long-period variables \citep[][black crosses]{mend67}.                    
     }
      \label{f_loops_noFe}
\end{figure}

\begin{figure*}[ht]
\centering
\includegraphics[width=\hsize]{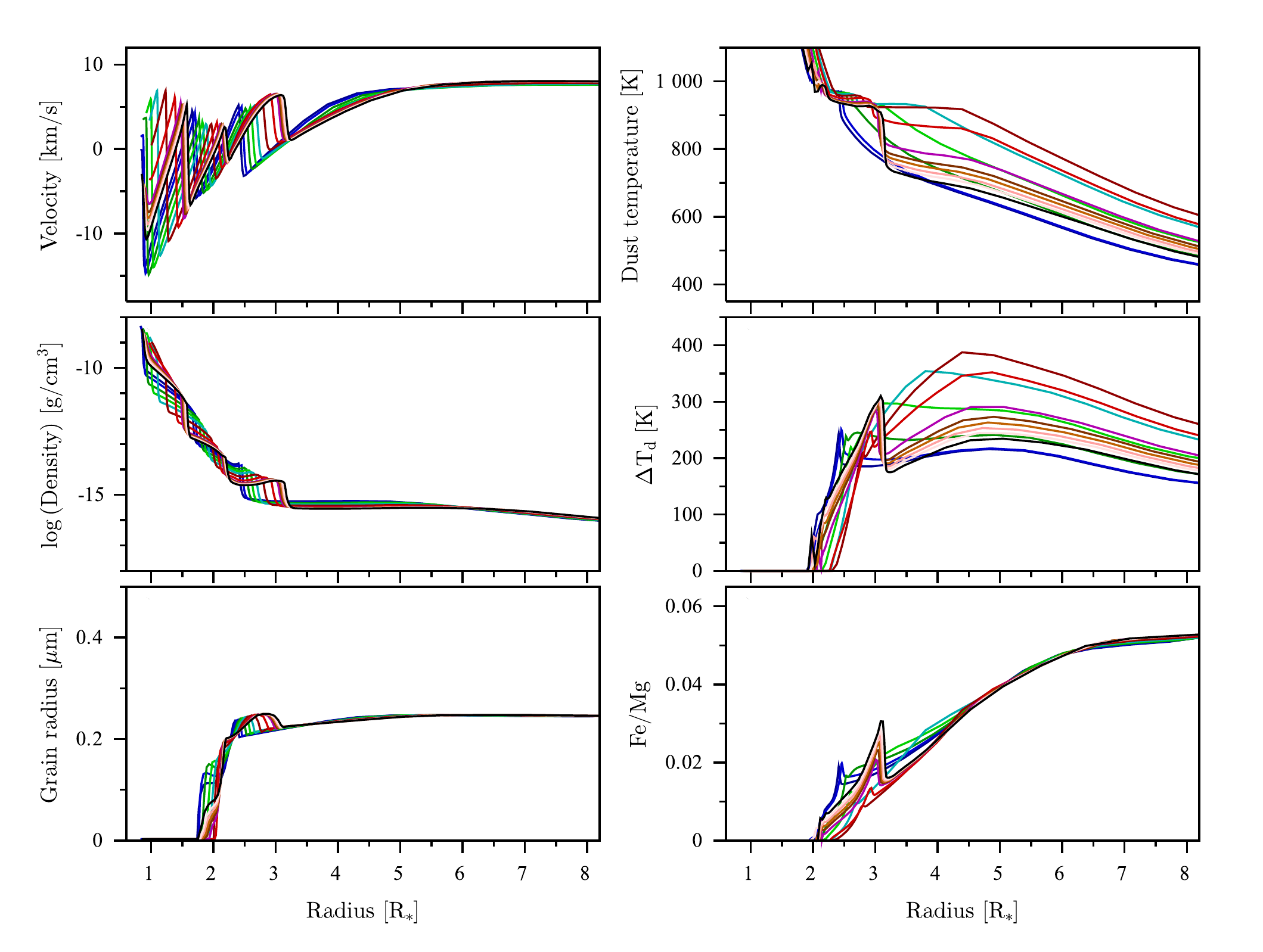}
     \caption{Time-dependent radial structure of model Bn114u4, zoomed in on the dust formation region (snapshots of 13 phases during a pulsation cycle).  
     {\em Left, top to bottom}: Flow velocity, gas density, grain radius.
     {\em Right, top to bottom}: Temperature of the Fe-bearing silicate grains, difference in grain temperature with and without Fe, and the Fe/Mg ratio in the dust grains; see text for details. The first snapshot, shown in dark blue, corresponds to a near-minimum phase, green colors represent the ascending part of the bolometric light curve, the red curves show phases close to the luminosity maximum, and the remaining colors represent the descending part of the bolometric light curve, ending with the black curve. We note that only the innermost, dust-free parts of the model structures show periodic variations that repeat every pulsation cycle (with the final black curve close to the initial dark blue curve in the velocity and density plots), while grain growth and wind acceleration are governed by other timescales. 
     }
      \label{f_struct2}
\end{figure*}

In Fig.~\ref{f_SEDs_noFe} we compare the SED of the new model An315u4 with Fe-bearing silicate grains (Table~\ref{t_mod}) to an earlier version based on Fe-free silicates, but with otherwise identical parameters \citep[Table~\ref{t_mod_noFe}; see also][]{hoefetal16}. As discussed above, the wind properties and the amount of dust in the two models are quite similar. Nevertheless, the model with Fe-free dust grains (consisting entirely of Mg$_2$SiO$_4$) does not show the characteristic silicate features around 10 and 18\,$\mu$m. This is a consequence of much lower dust temperatures for Fe-free grains, due to less radiative heating (lower absorption in the visual and near-IR spectral region). This effect has been discussed in some detail in previous papers \citep[see, e.g.,][]{bladetal15,bladetal17}. Figure~\ref{f_loops_noFe} shows how visual and near-IR colors of these two models vary during a pulsation cycle (similar to Fig.\,\ref{f_loops}). As expected, the loop traced out by the model with Fe-bearing silicate grains is shifted toward slightly redder values due to the higher dust absorption, especially in the V-band.  

In Sect.\,\ref{s_discussion} we propose that silicate condensation distances commonly derived from spectro-interferometric observations (typically $4-5\,\Rstar$) correspond to a region where Fe/Mg has reached a few percent. This makes the grains warm enough to produce emission features at mid-IR wavelengths, in contrast to the basically Fe-free silicate dust forming at about 2$\,\Rstar$ and initiating the wind. To demonstrate that the strong radiative heating of Fe-enriched silicate grains around $4-5\,\Rstar$ seen in in Fig.~\ref{f_struct} is not specific to that particular model, Fig.~\ref{f_struct2} shows a corresponding plot for a model with different stellar and pulsation parameters, resulting in different wind and dust properties. In this context it is interesting to note that 
\cite{sacuetal13} detected weak traces of a 10\,$\mu$m silicate feature at about $2\,\Rstar$ in RT Vir, that is, at a distance where the bulk of the wind-driving Fe-free silicate material has condensed in the DARWIN models but Fe enrichment is still very low. 

\end{appendix}


\end{document}